\documentclass[cernpreprint, atlasdraft=false, texlive=2023, UKenglish, coverpage=false, PAPER=true, texmf, orcidlogo]{atlasdoc}
 
\usepackage{atlaspackage}
\usepackage{floatrow}
\usepackage{atlasbiblatex}
 
\usepackage{atlasphysics}

\graphicspath{{logos/}{figures/}}
 
\usepackage{ANA-HIGG-2018-20-PAPER-defs}
\usepackage{multirow}

\AtlasTitle{Measurement of the $VH, H\rightarrow \tau\tau$ process with the ATLAS detector at 13 TeV}
 
\AtlasVersion{1.1}
 
\AtlasAbstract{
A measurement of the Standard Model Higgs boson produced in
association with a $W$ or $Z$ boson and decaying into a pair of
$\tau$-leptons is presented.
This search is based on proton-proton collision data collected at
$\sqrt{s}=13$\,~TeV by the ATLAS experiment at the LHC corresponding
to an integrated luminosity of 140 fb$^{-1}$. For the Higgs boson
candidate, only final states with at least one $\tau$-lepton decaying
hadronically ($\tau\rightarrow \mathrm{hadrons} + \nu_\tau$) are
considered. For the vector bosons, only leptonic decay channels are
considered: $Z \rightarrow \ell\ell$ and $W\rightarrow \ell\nu_\ell$,
with $\ell=e,\mu$.
An excess of events over the expected background is found with an
observed (expected) significance of 4.2 (3.6) standard deviations,
providing evidence of the Higgs boson produced in association with a
vector boson and decaying into a pair of $\tau$-leptons. The ratio of
the measured cross-section to the Standard Model prediction is
$\mu_{\text{VH}}^{\tau\tau} = 1.28\ ^{+0.30}_{-0.29}\
(\mathrm{stat.})\ ^{+0.25}_{-0.21}\ (\mathrm{syst.})$. This result
represents the most accurate measurement of the \vh process
achieved to date. }

\AtlasRefCode{HIGG-2018-20}
 
\PreprintIdNumber{CERN-EP-2023-272}

\AtlasJournal{Phys.\ Lett.\ B.}
\AtlasJournalRef{Phys. Lett. B 855 (2024) 138817}
\AtlasDOI{10.1016/j.physletb.2024.138817}

\hypersetup{pdftitle={ATLAS document},pdfauthor={The ATLAS Collaboration}}
 
\begin{document}
 
\maketitle
 
\tableofcontents

\section{Introduction}
 
This paper presents a search for the associated production of
the Higgs boson with a vector boson in which the Higgs boson decays to
a pair of $\tau$-leptons.  This process is referred to as $\vh$,
where $V$ represents either a $W$ or $Z$ boson. Two possible final
states are considered for the $H\rightarrow\tau^+\tau^-$ decay:
either both $\tau$-leptons decay hadronically to one or more hadrons
($\tau_{\text{had}}\tau_{\text{had}}$), or one $\tau$-lepton decays
leptonically ($\tau \rightarrow \ell \nu_\tau\bar{\nu}_\ell,
\ \ell=e,\mu$) and one hadronically
($\tau_{\text{lep}}\tau_{\text{had}}$).
The combination in which both $\tau$-leptons from the Higgs boson
decay leptonically ($\tau_{\text{lep}}\tau_{\text{lep}}$) is not
included in order to ensure an event selection that is independent
from analyses such as the one presented in
Ref.~\cite{HIGG-2017-14}. 
 
The events associated with the $\vh$ process are classified by the
leptonically-decaying vector boson candidate (either a $W$ or $Z$
boson) and by the Higgs boson decay channel
($\tau_{\text{had}}\tau_{\text{had}}$ or
$\tau_{\text{lep}}\tau_{\text{had}}$) into four channels. Each channel
is independently optimised to separate these rare events from their
background processes and to maximise the sensitivity of the analysis.
The vector boson decaying to light leptons provides an efficient
trigger option for these events that does not require relying on the
Higgs boson decay products.
Both the $VH$ production~\cite{HIGG-2018-04,CMS-HIG-18-016} and the
$H\rightarrow\tau\tau$ decay
channels~\cite{HIGG-2017-07,CMS-HIG-16-043,ATLAS:2022yrq} have separately been
observed by the ATLAS and CMS experiments in recent years. Recently,
the CMS collaboration measured a signal strength relative to the SM
prediction of the inclusive $\vh$ process to be $1.79\pm
0.45$~\cite{CMS-HIG-19-010}.
 
A similar search was performed by the ATLAS collaboration using 20.3
fb$^{-1}$ of LHC Run 1 data at
$\sqrt{s}=8$~TeV~\cite{HIGG-2014-01}. Because of the smaller size of
the dataset used in this search, it was only able to set a 95\%
confidence level (C.L.)  upper limit on the overall $\vh$ cross
section of 5.6 times the SM prediction. Compared with the Run 1
result, the analysis presented in this paper uses nearly seven times
the total integrated luminosity. This analysis also benefits from the
higher center-of-mass energy of Run 2 ($\sqrt{s}=13$~TeV), providing
an increase in the Higgs boson production cross
section~\cite{deFlorian:2016spz} of a factor of about 2, and from
improved physics object reconstruction and calibration.  The event
selection is expanded through the addition of several channels, and
the overall analysis strategy is improved via a neural network (NN)
discriminant that enhances the signal vs. background selection
efficiency.


\newcommand{\AtlasCoordFootnote}{
ATLAS uses a right-handed coordinate system with its origin at the nominal interaction point (IP)
in the centre of the detector and the \(z\)-axis along the beam pipe.
The \(x\)-axis points from the IP to the centre of the LHC ring,
and the \(y\)-axis points upwards.
Cylindrical coordinates \((r,\phi)\) are used in the transverse plane,
\(\phi\) being the azimuthal angle around the \(z\)-axis.
The pseudorapidity is defined in terms of the polar angle \(\theta\) as \(\eta = -\ln \tan(\theta/2)\).
Angular distance is measured in units of \(\Delta R \equiv \sqrt{(\Delta\eta)^{2} + (\Delta\phi)^{2}}\).}

\section{The ATLAS detector}
 
The ATLAS detector~\cite{PERF-2007-01} at the LHC covers nearly the
entire solid angle around the collision
point\footnote{\AtlasCoordFootnote}. It consists of an inner tracking
detector surrounded by a thin superconducting solenoid,
electromagnetic and hadron calorimeters, and a muon spectrometer
incorporating three large superconducting air-core toroidal magnets.
 
The inner-detector system (ID) is immersed in a \SI{2}{\tesla} axial
magnetic field and provides charged-particle tracking in the range
\(|\eta| < 2.5\).  The high-granularity silicon pixel detector covers
the vertex region and typically provides four measurements per track,
the first hit normally being in the insertable B-layer (IBL) installed
before Run~2~\cite{ATLAS-TDR-19,PIX-2018-001}.  It is followed by the
silicon microstrip tracker (SCT), which usually provides eight
measurements per track.  These silicon detectors are complemented by
the transition radiation tracker (TRT), which enables radially
extended track reconstruction up to \(|\eta| = 2.0\).  The TRT also
provides electron identification information based on the fraction of
hits (typically 30 in total) above a higher energy-deposit threshold
corresponding to transition radiation.
 
The calorimeter system covers the pseudorapidity range $(|\eta| <
4.9)$.  The solid angle coverage for $|\eta|$ between 3.2 and 4.9 is
completed with copper/liquid-argon (LAr) and tungsten/LAr calorimeter
modules optimised for electromagnetic and hadronic measurements,
respectively.
Within the region \(|\eta|< 3.2\), electromagnetic calorimetry is
provided by barrel and endcap high-granularity lead/LAr calorimeters,
with an additional thin LAr presampler covering \(|\eta| < 1.8\) to
correct for energy loss in material upstream of the calorimeters.
Hadron calorimetry is provided by the steel/scintillator-tile
calorimeter, segmented into three barrel structures within \(|\eta| <
1.7\), and two copper/LAr hadron endcap calorimeters.  The solid angle
coverage is completed with forward copper/LAr and tungsten/LAr
calorimeter modules optimised for electromagnetic and hadronic energy
measurements respectively.
 
The muon spectrometer (MS) comprises separate trigger and
high-precision tracking chambers measuring the deflection of muons in
the magnetic field generated by the superconducting air-core toroidal
magnets. The field integral of the toroids ranges between \num{2.0}
and \SI{6.0}{\tesla\metre} across most of the detector.  Three layers
of precision chambers, each consisting of layers of monitored drift
tubes, cover the region \(|\eta| < 2.7\), complemented by
cathode-strip chambers in the forward region, where the background is
highest.  The muon trigger system covers the range \(|\eta| < 2.4\)
with resistive-plate chambers in the barrel, and thin-gap chambers in
the endcap regions.
 
Interesting events are selected by the first-level trigger system
implemented in custom hardware, followed by selections made by
algorithms implemented in software in the high-level
trigger~\cite{TRIG-2016-01}.  The first-level trigger accepts events
from the \SI{40}{\MHz} bunch crossings at a rate below \SI{100}{\kHz},
which the high-level trigger further reduces in order to record events
to disk at about \SI{1}{\kHz}.
 
An extensive software suite~\cite{ATL-SOFT-PUB-2021-001} is used in
the reconstruction and analysis of real and simulated data, in
detector operations, and in the trigger and data acquisition systems
of the experiment.


\section{Data and simulation samples}
The dataset used for this measurement consists of the LHC
proton--proton collision data recorded by the ATLAS experiment at
$\sqrt{s}=13$~\TeV{} during the Run 2 period from 2015 to 2018. Events are
selected for analysis only if they are of good quality and if
all the relevant detector components are known to have been in good
operating condition~\cite{DAPR-2018-01}. The total integrated
luminosity of the analysed data is \ilumi.
Only events that pass relevant trigger requirements are considered in
the analysis. These triggers are designed to select single electrons,
single muons, or combinations of these two light
leptons~\cite{TRIG-2018-05,TRIG-2018-01,TRIG-2019-03,TRIG-2019-02}. The
thresholds applied to the reconstructed transverse momentum (\pt)
for the single--lepton triggers were $\pt^{e} >
27\,(25)$~\GeV\ and $\pt^{\mu} > 27.3\,(21)$~\GeV\ for the 2016--2018
(2015) data-taking period. The \pt{} thresholds for the
dilepton triggers were $\pt^{e} > 18$~\GeV\ and $\pt^{\mu} >
14.7$~\GeV for the entire data-taking period.
For the $ZH$ events, the trigger selection acceptance was
maximized by using the logical OR of triggers requiring one or two
light leptons.
 
Samples of Monte Carlo (MC) simulated events are used to optimise the
event selection and to model the signal and several background
processes.
Simulated event samples for the $\vh$ signal, as well as all
background samples, were produced using various MC generators, as
described in Table \ref{tab:MCgenerators}.
The samples were produced with the ATLAS simulation
infrastructure~\cite{SOFT-2010-01} using the full detector simulation
performed by the {\GEANT}~\cite{Agostinelli:2002hh} toolkit.  The
\textsc{Powheg} NNLOPS
program~\cite{Frixione:2007nw,Nason:2004rx,Frixione:2007vw,Alioli:2010xd,Hartanto:2015uka}
was used to model gluon-gluon fusion (ggF) Higgs boson production with
next-to-next-to-leading-order (NNLO) accuracy with the
\textsc{PDF4LHC15NLO}~\cite{Butterworth:2015oua} parton distribution
function (PDF) set. The vector boson fusion (VBF) and the $VH$
production processes were simulated with \textsc{Powheg} at
next-to-leading-order (NLO) accuracy in QCD using the
\textsc{PDF4LHC15NLO}~\cite{Butterworth:2015oua} PDF set.
The MC prediction of the Higgs production modes mentioned above was
normalized to cross-sections calculated at NNLO in the strong coupling
with NLO electroweak
corrections~\cite{Ciccolini:2003jy,Brein:2003wg,Brein:2011vx,Denner:2014cla,Brein:2012ne}.
The production of $\ttH$ events was simulated using
\textsc{PowhegBox}~\cite{Frixione:2007nw,Nason:2004rx,Frixione:2007vw,Alioli:2010xd,Hartanto:2015uka}
at NLO using the \textsc{NNPDF30NNLO}~\cite{Ball:2012cx} PDF set.
In all signal events, the decays of the $\tau$-leptons were modelled
by \textsc{Pythia8.235}~\cite{Sjostrand:2014zea}. Background samples
of \Vjets use \textsc{Sherpa 2.2.1}~\cite{Bothmann:2019yzt} with NNLO
accuracy and \textsc{NNPDF30NNLO}~\cite{Ball:2012cx} PDF, while the diboson and triboson events were generated by
\textsc{Sherpa 2.2.2}~\cite{Bothmann:2019yzt} (including $\tau$-lepton
decays) at NNLO with \textsc{NNPDF30NNLO}~\cite{Ball:2012cx} PDF, and $\ttbar$ and single--top samples were generated
by \textsc{Powheg+Pythia8.230} with NLO accuracy using the \textsc{NNPDF30NLO}~\cite{Ball:2012cx} PDF, with \textsc{Pythia}
also performing $\tau$-lepton decays.
The \textsc{Powheg+Pythia8} samples use \textsc{EvtGen(v1.6.0)}~\cite{Lange:2001uf}
for the simulation of the $b$-hadron decays.
 
The effects of multiple interactions in the same and neighbouring
bunch crossings (pile-up) were modeled by overlaying minimum-bias
events to reproduce the pile-up distributions seen in the data. These
minimum-bias events were simulated using the soft QCD processes of
\PYTHIA[8.186]~\cite{Pythia8} with the A3~\cite{ATL-PHYS-PUB-2016-017}
set of tuned parameters and the \NNPDF[2.3lo]~\cite{Ball:2012cx} PDF.
 
\begin{table}[h]
\centering
\begin{tabular}{lccc}
Process & MC Generator + UEPS & PDF Set & Perturbative Order \\
\hline
\multicolumn{4}{l}{Signal}\\
\hline
$W\rightarrow \ell\nu$, $H\rightarrow\tau\tau$ & \textsc{Powheg~\cite{Frixione:2007nw,Nason:2004rx,Frixione:2007vw,Alioli:2010xd,Hartanto:2015uka}+Pythia8.235~\cite{Sjostrand:2014zea}} & \textsc{PDF4LHC15NLO}~\cite{Butterworth:2015oua} & \textsc{NLO} \\
$Z\rightarrow \ell\ell$, $H\rightarrow\tau\tau$ & \textsc{Powheg+Pythia8.235} & \textsc{PDF4LHC15NLO}  & \textsc{NLO} \\
\hline
\multicolumn{4}{l}{Background}\\
\hline
ggF $H\rightarrow\tau\tau$                    & \textsc{Powheg+Pythia8.235} & \textsc{PDF4LHC15NLO} & N\textsc{NLO} \\
VBF $H\rightarrow\tau\tau$                      & \textsc{Powheg+Pythia8.235} & \textsc{PDF4LHC15NLO}                      & \textsc{NLO} \\
$\ttH$, $H\rightarrow\tau\tau$                  & \textsc{Powheg+Pythia8.235} & \textsc{NNPDF30NNLO}~\cite{Ball:2012cx}    & \textsc{NNLO} \\
Diboson  & \textsc{Sherpa 2.2.2}~\cite{Bothmann:2019yzt} & \textsc{NNPDF30NNLO} & \textsc{NNLO} \\
Triboson & \textsc{Sherpa 2.2.2}                         & \textsc{NNPDF30NNLO} & \textsc{NNLO} \\
\Vjets   & \textsc{Sherpa 2.2.1}~\cite{Bothmann:2019yzt} & \textsc{NNPDF30NNLO} & \textsc{NNLO} \\
Single--top & \textsc{Powheg+Pythia8.230}                     & \textsc{NNPDF30NLO}  & \textsc{NLO}  \\
$\ttbar$   & \textsc{Powheg+Pythia8.230}                     & \textsc{NNPDF30NLO}  & \textsc{NLO}  \\
 
\end{tabular}
\caption{Information on the Monte Carlo event samples used to
produce the most relevant processes incorporated into this
analysis, including the process name, names of the MC generator
and the model of the underlying event with hadronisation and
parton showering (UEPS), the corresponding PDF set, and the
perturbative order in QCD to which the events were generated.}
\label{tab:MCgenerators}
\end{table}


\section{Object reconstruction and event selection}
The \vh{} event selection requires the reconstruction of electrons,
muons, visible products of hadronically decaying $\tau$-leptons
(\tauhadvis), jets (along with their flavour tagging properties),
and missing transverse energy (\met{}). The number of reconstructed
electrons, muons and \tauhadvis{} in each event is used to separate
the events into analysis channels.

\subsection{Object reconstruction}
\label{sec:objects}

Electron candidates are reconstructed from tracks in the inner
detector matching calorimeter energy deposits~\cite{EGAM-2021-01}. The
electron candidates must fulfill the following baseline requirements:
$\pt>13$~GeV, a pseudorapidity $|\eta|$ below 2.5 and not in the
barrel-endcap transition region ($1.37 <|\eta|<1.52$), and passing the
Loose likelihood selection requirement (93\% efficient) for electron
identification~\cite{EGAM-2021-01}.  For an electron to qualify for
one of the signal-enhanced categories, it must additionally pass the
Tight identification selection requirement (80\% efficiency) and the
Loose isolation criterion, which is defined for both calorimeter and
track-based isolation~\cite{EGAM-2021-01}. 
level, the
 
Muon candidates are reconstructed from tracks in the muon spectrometer and then
matched to tracks in the inner detector~\cite{MUON-2018-03}.  Baseline
muon candidates included in the analysis are required to pass a minimum $\pt$
threshold of 9~GeV, have an $|\eta| < 2.5$, and pass the Loose muon
identification selection requirement (corresponding to over 98\%
efficiency).
For a muon to qualify for one of the signal-enhanced
categories, it must pass the Tight selection requirement (with an
efficiency between 90 and 93\%, depending on the $\pt$ of the
muon). Selected muon candidates must also pass a Tight
isolation criterion that is based exclusively on tracking
information.~\cite{MUON-2018-03}

Jets are reconstructed from particle flow objects using the
anti-$k_t$~\cite{Cacciari:2008gp,Fastjet} algorithm with a
distance parameter $R=0.4$, and calibrated as in
Ref.~\cite{JETM-2018-05}.
Additional requirements on the jet-vertex-tagger
(JVT)~\cite{PERF-2014-03} are imposed to suppress jets originating
from pile-up.
In order to identify jets initiated by \emph{b} quarks for suppression
of top quark backgrounds in $WH(\tau\tau)$ events, the DL1r
$b$--tagging
algorithm~\cite{ATLAS-CONF-2018-045,FTAG-2018-01,ATL-PHYS-PUB-2017-013,FTAG-2019-07}
is used on jets with $\pt>20$~GeV and $|\eta|<2.5$. The fixed 85\%
efficiency selection requirement is inverted to reject $b$-jets. The
rejection factors for $b$-tagged jets initiated by $c$-quarks and
light partons are 2.6 and 29 respectively for the 85\% efficiency
working point.

The final states of $\tau$-lepton hadronic decays include a neutrino and a set of
visible decay products, most frequently one or three charged pions and
up to two neutral pions.  The reconstruction of the \tauhadvis{} is
seeded by jets reconstructed via the anti-$k_t$ algorithm, using
calibrated energy clusters as inputs, with a distance parameter
of $R = 0.4$~\cite{ATL-PHYS-PUB-2015-045}.  Jets seeding \tauhadvis{}
candidates are additionally required to have $\pt > 10\,\GeV$~ and
$|\eta|< 2.5$.  To separate the \tauhadvis{} candidates initiated by
hadronic $\tau$-lepton decays from jets initiated by quarks or gluons, a
Recurrent Neural Network (RNN)~\cite{ATL-PHYS-PUB-2019-033}
identification method was trained on information from
reconstructed charged-particle tracks and clusters of energy in the
calorimeter associated to \tauhadvis{} candidates as well as
high-level discriminating variables.
A separate multivariate discriminant based on a Boosted Decision
Tree~\cite{ATL-PHYS-PUB-2022-044} is also used to reject backgrounds
arising from electrons mimicking a \tauhadvis{} (mainly from \Zeejets{} in
this analysis).  This discriminant (\textsc{eBDT}) is built using
information from the calorimeter and the tracking detector.
Transition radiation information from the TRT system plays a key role
in the performance of this discriminant.  Baseline \tauhadvis{} are required
to have 1 or 3 associated tracks, electric charge of $\pm1$, $\pt >
20\,\GeV$ and $|\eta|< 2.5$, excluding the barrel-endcap transition region.
In addition, a dedicated muon veto criterion, designed to reject muons
misreconstructed as taus (typically due to large calorimeter energy
deposits), is applied.
To qualify as objects at the earliest stage of the analysis event
categorisation, each \tauhadvis{} must pass the Medium RNN
identification selection requirement, with efficiencies of 75\% for
1-prong and 60\% for 3-prong taus, and the Loose \textsc{eBDT}
requirement, with an efficiency of 95\%.\footnote{For the \whlephadshort{} category, as defined
in Section~\ref{sec:evt_cat}, the Medium \textsc{eBDT} requirement is
used. This has an efficiency of 85\%.}~\cite{ATL-PHYS-PUB-2022-044}.

An overlap procedure is applied to ensure that electrons, muons,
\tauhadvis{} and jets used in this analysis are built from a set of
mutually exclusive tracks and calorimeter energy deposits.
More details can be found in Ref.~\cite{HIGG-2019-09}.

The missing transverse momentum vector is an estimate of the imbalance in the
transverse momentum in the detector. This vector is calculated as the
negative vector sum of the transverse momenta of all reconstructed
final-state objects (electrons, muons, taus, and jets). The magnitude
of the missing transverse momentum vector is referred to as missing transverse
momentum (\met{}).  Tracks not associated to any reconstructed object
are also included in the calculation, and serve to estimate the
contribution from low-\pt{} collision remnants (referred to as the
\emph{soft term} contribution).
The default Tight criterion of the official ATLAS Missing Transverse
Energy Tool was chosen~\cite{ATLAS-CONF-2018-023}. Tight \met{} is
calculated without including forward jets with $|\eta| > 2.4$ and $20
< \pt < 30$~GeV. This tighter threshold removes regions of phase space
that have more pileup jets than hard scatter jets.


\subsection{Event categorization and selection}\label{sec:evt_cat}
The analysis channels are defined by the vector boson
associated with the Higgs boson production and by the decay mode
(leptonic or hadronic) of the $\tau$-leptons associated with the Higgs
boson decay.
This results in four different channels: \whlephadshort,
\whhadhadshort, \zhlephadshort{} and \zhhadhadshort. In all channels,
only the final states in which the vector boson decays to light
leptons are considered. In the \alephad{} channel, the leading \pt{}
light lepton is assigned to the vector boson ($V$). If $V$ is a $Z$
boson, the other light lepton assigned to the $Z$ boson leptonic decay
is required to have the same flavor and opposite electric charge
(referred to as ``sign'' throughout the paper) to that of the leading
\pt{} light lepton. If multiple opposite-sign light lepton pairs can
be formed, the pair with invariant mass closest to the $Z$ boson mass
is chosen. The same invariant mass is also required to be within a
given mass range, optimized separately for each category, to suppress
events with a pair of light leptons not originating from a \Zll{}
process. The light lepton not assigned to the $V$ boson leptonic decay
is then associated to the leptonic decay of the $\tau$-lepton from the
Higgs boson. When $V$ is a $W$ boson, the sub-leading \pt{} light
lepton is assigned to the Higgs boson with no flavor or charge
selection applied.
For each category, the selection is organized into a
\textsc{Preselection} and \textsc{Signal Region} selection. The
\textsc{Preselection} is used as a starting point of shared selections from which a series
of validation regions are further defined to check the modelling of specific background
sources.
The light lepton from the \taulep{} decay in the \whlephadshort{} and
\zhlephadshort{} \textsc{Signal Regions} events has a small probability
of about 3\% to be misassigned to the vector boson instead of the
Higgs boson because of the kinematic selections applied.
 
The main analysis strategy uses a NN classifier as a final
discriminant. As a cross-check, another version of the analysis is
performed using Higgs boson mass estimators as the final discriminants, as
was done in the Run 1 analysis~\cite{HIGG-2014-01}. In the $ZH$
channels, the Missing Mass Calculator (\mmmc)~\cite{MMCpaper} is used
to estimate the Higgs boson mass.
The \mmmc{} method is a precise strategy for calculating a
most likely parent particle mass when that parent particle decays into
multiple sources of \met{}, as in most of the $H\rightarrow\tau\tau$
event topologies.
For the $WH$ channels, in which both the $W$ and the Higgs boson decays
act as sources of \met{}, the \mmmc{} assumption that the \met{} (i.e. neutrinos)
originate only from the Higgs boson decay is no longer valid, so the
Late-Projected transverse mass $\mtwot$~\cite{M2Tpaper} is instead used as a
discriminant\footnote{The $\mtwot$ variable is constructed to provide
an event--by--event lower bound on the transverse mass of the heaviest
parent particle -- in this topology, the Higgs boson.
The \mtwot{} is defined as $\mtwot=\min_{\sum\vec{q}_{iT}=\ptmiss}
\left[ \max_{a}[m_{aT}]\right]$, where $m_{aT}$ is the late-projected
transverse mass of the $a^{\ensuremath{th}}$ parent particle and
$\sum\vec{q}_{iT}$ is the sum of the transverse momenta of all
invisible particles.}.
 
Table~\ref{table:selections} summarizes the selection criteria used
for each category. The event selection shown in
Table~\ref{table:selections} is identical for the cross-check analysis using the Higgs boson mass discriminants, with the
exception of the Higgs boson mass window cuts. The combined signal
efficiency is about 6\% and 8\% for the $WH$ and $ZH$ channels,
respectively\footnote{Estimated using MC simulations and normalized to
the total number of signal events from a specific process falling into
the detector acceptance with no trigger requirements.}.
The \whlephadshort{} selection has a signal efficiency of about 7\%
and represents about 44\% of all the \vh{} signal events across the
four categories.
The \whhadhadshort{} selection has a signal efficiency of about 5\%
and represents about 32\% of all the \vh{} signal events across the
four categories.
The \zhlephadshort{} selection has a signal efficiency of about 7\% and
represents about 10\% of all the \vh{} signal events across the four
categories.
The \zhhadhadshort{} selection has a signal efficiency of about 9\% and
represents about 14\% of all the \vh{} signal events across the four
categories.

\begin{table}[htb!]
\centering
\resizebox{\textwidth}{!}{
 
\begin{tabular}{l|c|c|c|c}
\toprule
Selection                     &                 \whlephad                     &         \whhadhad         &       \zhlephad          &  \zhhadhad \\
\midrule
\multirow{4}{*}{\textsc{Preselection}}  & exactly 1 \tauhadvis{}  & exactly 2 \tauhadvis{}  & exactly 1 \tauhadvis{}         & exactly 2 \tauhadvis{}\\
& exactly 2 $\ell$                        & exactly 1 $\ell$                      & exactly 3 $\ell$                       & exactly 2 $\ell$             \\
&   $b$-jet veto                                & $b$-jet veto                                & same-flavour, OS $\ell$ pair                 & same-flavour, OS $\ell$ pair\\
&                                               &                                             & $\mathrm{m}_{\ell\ell} \in [81,101]$ GeV               & $\mathrm{m}_{\ell\ell} \in [71,111]$ GeV\\
\midrule
\multirow{4}{*}{\textsc{Signal Region}} &  1 \tauhadvis{} and 1 $\tau_{\text{lep}}$ OS  & exactly 2 \tauhadvis{} OS                   & exactly 1 \tauhadvis{} and 1 $\taulep$ OS    & exactly 2 \tauhadvis{} OS\\
& exactly 2 $\ell$  SS      & $0.8 < \Delta R(\tauhadvis,\tauhadvis) < 2.8$  & $\sum_{\tauhadvis,\taulep} \pT(\tau) > 60$ GeV& $\sum_{\tauhadvis} \pT(\tau) > 75$ GeV\\
& $\sum_\ell \pT(\ell) + \pT(\tauhadvis) > 90$ GeV          & $\sum_{\tauhadvis} \pT(\tauhadvis) > 100$ GeV            & & \\
& $\mathrm{m}_{ee}\notin [80,100]$ GeV                   & $\mathrm{m}_T(\ell,E_T^{\text{miss}}) > 20$ GeV        &  & \\
\midrule
\multirow{1}{*}{\textsc{Higgs boson mass window cut}}        & \multirow{2}{*}{$\mtwot \in [60,130]$ GeV} & \multirow{2}{*}{$\mtwot \in [80,130]$ GeV}            &   \multirow{2}{*}{ $\mmmc\in[100,170]$ GeV} & \multirow{2}{*}{$\mmmc\in[100,180]$ GeV} \\
\multirow{1}{*}{\textsc{(only applied in the NN-based analysis)}} &                                            &                                                                  &     &  \\

\bottomrule
\end{tabular}}
\caption{\textsc{Preselection} and \textsc{Signal Region} selection
for the four categories. ``OS'' stands for opposite-sign, ``SS'' for
same-sign. }
 
\label{table:selections}
 
\end{table}
 
\normalsize
\subsubsection{Selection of the \whlephad{} events}
The \textsc{Preselection} begins with requiring exactly two light
leptons and one \tauhadvis{} that pass their baseline requirements as
well as additional isolation and identification criteria as described
in Section \ref{sec:objects}. A $b$-jet veto is applied to suppress
backgrounds from top quark production.

For the \textsc{Signal Region} selection, a same-sign light lepton
requirement is applied to suppress \Zjets{} backgrounds and the
\tauhadvis{} is required to have the opposite sign of the light
leptons. A differentiation based on the final state light leptons is
also used to further optimise the selection criteria. For the
dielectron final state, the invariant mass of the two electrons
$\mathrm{m}_{ee}$ must satisfy $\mathrm{m}_{ee}\notin[80,100]$~GeV to reduce the
contamination of \Zjets{} events with $Z\rightarrow ee$ where one of
the $e$ was reconstructed with the wrong charge, thereby passing the
same-sign light lepton requirement. The scalar sum of all three final
state particles' $\pt$ is required to be greater than 90~GeV to
suppress events that contain one or two misidentified objects. Finally, for the main analysis strategy using a NN
discriminant, the $\mtwot$ is required to fall within $60 < \mtwot<
130$~GeV to improve the signal-to-background ratio.

\subsubsection{Selection of the \whhadhad{} events}
The \textsc{Preselection} begins with requiring exactly one
light lepton and two \tauhadvis{} that pass their baseline
requirements as well as additional isolation and identification
criteria to qualify as analysis-level objects, as described in Section
\ref{sec:objects}.  A $b$-jet veto is applied to suppress backgrounds
from top-quark production.
 
The \textsc{Signal Region} selection adds the following criteria. The
two \tauhadvis{} are required to have opposite sign. A
requirement on the radial distance between the two \tauhadvis{}
candidates ($0.8 < \Delta R(\thadvis,\ \thadvis) <2.8$) and on the
scalar sum of the $\pt$ of the two (\tauhadvis{} > 100~GeV) are
imposed to suppress events that contain one or two \fake{} passing the
selection. An additional cut on the transverse mass\footnote{The
transverse mass variable is defined by projecting the momenta to the
plane perpendicular to the beam direction: $\mt^2 =
\left(\sum{E_{iT}}\right)^2 - \left(\sum{\vec{p}_{iT}}\right)^2$, with
$E_{iT} = \sqrt{m_i^2 + \vec{p}_{iT}^2}$.}  between the \met{} and the
light lepton $\mt(\ell,\met)> 20$~GeV is applied to reduce the $\Ztt$
background. Finally, for the main analysis strategy using a NN
discriminant, the $\mtwot$ is required to fall within $80 < \mtwot <
130$~GeV to improve the signal-to-background ratio.

\subsubsection{Selection of the \zhlephad{} events} 
The \textsc{Preselection} begins with requiring exactly three light
leptons and one \tauhadvis{} that pass their baseline requirements as
well as additional isolation and identification criteria to qualify as
analysis-level objects, as described in Section \ref{sec:objects}. Two
light leptons must have the same flavour and opposite signs, as they
are associated with the $Z$ boson decay. The $\mmmc$ needs to have
successfully converged.

The \textsc{Signal Region} selection further requires that the decay
products of the Higgs boson have opposite sign. The invariant mass of
the light leptons m$_{\ell\ell}$ must satisfy
$81<\mathrm{m}_{\ell\ell}<101$~GeV.
The scalar sum of $\pt$ from the two objects associated with the Higgs
boson decay is required to be greater than 60~GeV. Lastly, for the
main analysis strategy using a NN discriminant, the $\mmmc$ must
satisfy $100<\mmmc<170$~GeV to enhance the signal-to-background ratio.

\subsubsection{Selection of the \zhhadhad{} events} 
The \textsc{Preselection} begins with requiring exactly two
light leptons and two \tauhadvis{} that pass their baseline
requirements as well as additional isolation and identification
criteria to qualify as analysis-level objects, as described in
Section~\ref{sec:objects}. These two light leptons associated with the
$Z$ boson decay need to be same flavour and opposite-sign. The $\mmmc$ is required to have successfully converged.

For the \textsc{Signal Region} selection, the two \tauhadvis{} from
the Higgs boson decay are required to have opposite signs.  The invariant
mass of the light leptons $\mathrm{m}_{\ell\ell}$ must satisfy $71<
\mathrm{m}_{\ell\ell}<111$~GeV.
In this case, the $\mathrm{m}_{\ell\ell}$ acceptance window is
slightly larger than the \zhlephadshort{} category due to lower
background levels.
To enhance the suppression of the \fake{} events, the scalar sum of
the $\pt$ of the taus is required to be greater than 75~GeV. Lastly,
for the main analysis strategy using a NN discriminant, the \mmmc{}
must satisfy $100<\mmmc<180$~GeV.

\section{Background estimation}\label{sec:background}
The main backgrounds for this analysis consist of $VZ$ events and
\Zjets\ events in which a jet is misidentified as a light lepton
($\ell = {e,\mu}$) or \tauhadvis{}. The background contribution from
events with jets misidentified is estimated using a data-driven
technique. Other background components such as top quark decays,
$\ttH$, and $WWV$ triboson events are estimated using MC simulations.
 
In the $ZH$ categories, the $ZZ$ events are the main background source
and are estimated using simulated events. In both $ZH$ categories, the
$ZZ$ events form $\sim60\%$ of the total background, while the
\fakebkg{} events account for much of the remaining $\sim40\%$. The
other background sources (top-quark decays, triboson and $\ttH$
events) account for less than $1\%$ of the total background and are
estimated using simulations.
 
In the $WH$ categories, \fakebkg{} events represent $\sim70\%$ of the
total background. In both $WH$ categories, the $WZ$ events account
for $\sim30\%$ of the total background and are estimated using
simulated events. The other background sources (top quark decays,
$WWW$, and $\ttH$ events) form less than $2\%$ of the total
background and are also estimated using simulated events.
 
Due to the difficulties of validating the diboson background in
dedicated validation regions, the analysis relies on previous
measurements with higher statistics
available~\cite{STDM-2018-03,STDM-2016-15} and normalises this
background to the Standard Model expectation. No dedicated validation
region is used to extract the diboson background normalization from
the data.
 
The background from misidentified jets is evaluated using the \ff{}
method~\cite{HIGG-2013-32,HIGG-2013-21}. A similar background
estimation was used in the previous version of this
analysis~\cite{HIGG-2014-01}. A \ff{} is defined as $f = r/(1-r)$,
where $r$ represents the selection efficiency of misidentified objects
(\tauhadvis{} or light lepton). This means that $f$ is the ratio of
the number of objects passing the selection requirements (as described
in Section~\ref{sec:objects}) to the number of objects that pass
nearly all the selection requirements, but fail one or both of the
identification and isolation requirements. For electrons and muons, at
least one of the identification and isolation requirements must fail,
while $\tau$-leptons need to pass the Very-Loose identification
requirement (efficiency of 99\%) and fail the Medium identification
requirement.
The derived $r$ values range from 0.32 (0.07) at $\pT<25$~GeV to
0.02 (0.01) at $\pT>60$~GeV for 1-track (3-track) \tauhadvis{}
candidates. For the muon (electron) case, the $r$ values range from
0.14 (0.13) at $\pT<15 \ (10)$~GeV to 0.25 (0.36) at $\pT>20$~GeV.
 
The expected number of misidentified jets in a given region is
obtained using the \ff{} to scale the number of events selected in an
orthogonal region in which one or more requirements are inverted: the
identification and/or isolation requirements for light leptons, and
the identification requirements for the \tauhadvis.
The \ffs{} are measured in a dedicated $\Zjets$ control region (CR)
enriched in \fakebkg{}, and are then used as extrapolation factors to
estimate the number of selected fake objects in the signal
region. This $\Zjets$ CR is based on the selection of exactly two
opposite-sign light leptons that are required to be consistent with a
$Z$-boson decay, plus a third object that is assumed to originate from
a jet that is misidentified either as an electron, muon or
\tauhadvis{} and is used for the determination of the corresponding
\ffs{}.
The \ffs{} are computed in bins of lepton \pT{} and $|\eta|$. The
\tauhadvis{} \ffs{} are also parametrized with respect to the
\tauhadvis{} jet width, defined as a weighted sum of jet constituent
distance from the jet axis ($\mathrm{jet\ width} = \sum_i \Delta R^i
\pT^i / \sum_i \pT^i$) separately for one- and three-track
\tauhadvis{} candidates.  In the \tauhadvis{} case, the jet width
parametrization is particularly important because it is highly
correlated with the jet quark--gluon fraction composition, and quark-
vs. gluon-initiated jets exhibit different \tauhadvis{}
misidentification rates. Contributions from a jet misidentified as
a light lepton that triggered the event are estimated from the
simulation and found to be negligible.
Diboson events can contaminate the selection at a level below a few
percent, and are therefore subtracted using the MC prediction.
The evaluation of the \fakebkg{} takes into account the presence of
multiple misidentified objects, which can be as high as three in the
$ZH$ categories in which only one light lepton triggered the event.

In all the categories, the modelling of the \fakebkg{} is validated in
a misidentified background-enriched same-sign region. For the $ZH$
categories, the \textsc{Preselection} criteria are applied, and the
objects associated to the Higgs boson decay are required to have the
same charge. For the $WH$ categories, all the SR cuts are applied
except the \mtwot{} mass cut and same-sign $\tau$-leptons selection
requirement. The selected region contains a sufficiently large number
of events to minimize statistical fluctuations. Validation regions in
the $WH$ categories contain sufficient events to be able to verify the
modelling of the background from misidentified jets, as summarized in
Table~\ref{table:vrs}.The same table also shows the expected
composition of the processes contributing to the \fakebkg{} in each
region obtained entirely from simulation.
 
The uncertainties associated with the \ff method have statistical and
systematic components.  The statistical component is estimated for
each \ff bin separately and consists of the statistical uncertainties
from the data within the $\Zjets$ CR propagated to the \ffs. For each
category, a dedicated systematic uncertainty takes into account the
statistical fluctuations associated with the subtracted MC
component. Another systematic uncertainty accounts for the residual
difference between the \fakebkg{} modelling and the data in the
misidentified background-enriched same-sign region. This uncertainty
is evaluated at the \textsc{Preselection} stage, which requires the
objects associated to the Higgs boson decay to have the same
charge. For each category, this uncertainty is evaluated in bins of
\tauhadvis{} \pT{}.
The statistical component is negligible ($< 5\%$)
compared with the systematic uncertainty, which ranges up to 23\% in the
high \pT{} region (above 60~GeV).
 
\begin{table}[t]
\centering
\resizebox{\textwidth}{!}{
\begin{tabular}{l|c|c|c}
\toprule
Category                  & Region                    &                 Cuts                  &  Major process contributing      \\
&                           &                                       &  to the \fakebkg{}                 \\
 
\midrule
\multirow{8}{*}{\whhadhad} & \multirow{3}{*}{\Wjets{}} & \textsc{Preselection}                        &  \multirow{3}{*}{$\Wjets\sim70\%$}\\
&                           & same-sign \tauhadvis                     &  \\
&                           & $\mT(\ell,\MET) < 60$ GeV                &  \\
\cline{2-4}
& \multirow{3}{*}{\Ztt{}}   & \textsc{Preselection}                        &  \multirow{3}{*}{$\Ztt\sim50\%$}\\
&                           & $\mtwot<60$ GeV                       &  \\
&                           & $\mT(\ell,\MET) < 40$ GeV                &  \\
\cline{2-4}
& \multirow{2}{*}{top-quark}      & \textsc{Preselection}                        &  \multirow{2}{*}{$\ttbar\sim70\%$}\\
&                           & \# $b$ jets $>0$                         &  \\
\midrule
\multirow{7}{*}{\whlephad} & \multirow{4}{*}{\Ztt{}}   & \textsc{Preselection}                        &  \multirow{4}{*}{$\Ztt\sim40\%$}\\ 
&                           & opposite-sign light leptons           &  \\
&                           & $\mcol(\ell,\ell)\in [60,120]$ GeV    &  \\
&                           & $m_{ee}\notin [80,100]$ GeV           &  \\
\cline{2-4}
&\multirow{3}{*}{All Same Sign} & \textsc{Preselection}                  &  \multirow{3}{*}{$\Wjets\sim70\%$}\\
&                                 & all objects with same-sign      &  \\
&                                 & $m_{ee}\notin [80,100]$ GeV     &  \\
 
\bottomrule
\end{tabular}}
\caption{Validation regions, in addition to the misidentified
background-enriched same-sign region, used to check the modelling of
the \fakebkg{} in the $WH$ categories. The last column shows an
estimate of the major contribution to the \fakebkg, which was
estimated using a pure MC study. The definition of the colinear mass
($\mcol$) can be found in Ref.~\cite{Ellis:1987xu}.}
\label{table:vrs}
 
\end{table}
 
\begin{figure}[!htb]
\centering
\subfloat[]{\includegraphics[width=0.43\textwidth]{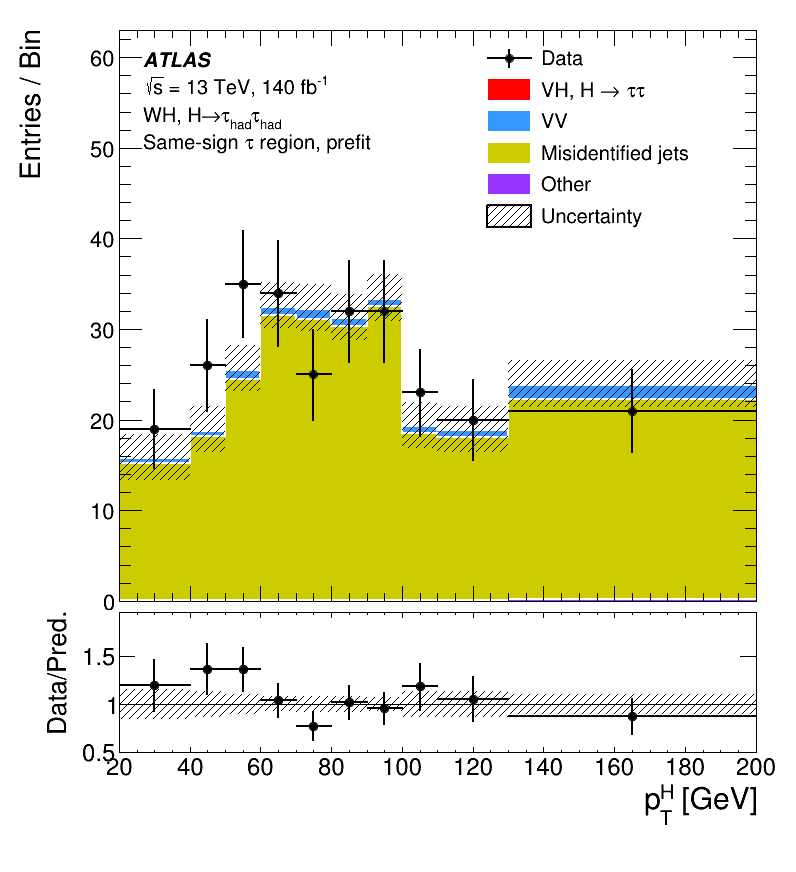}}
\subfloat[]{\includegraphics[width=0.43\textwidth]{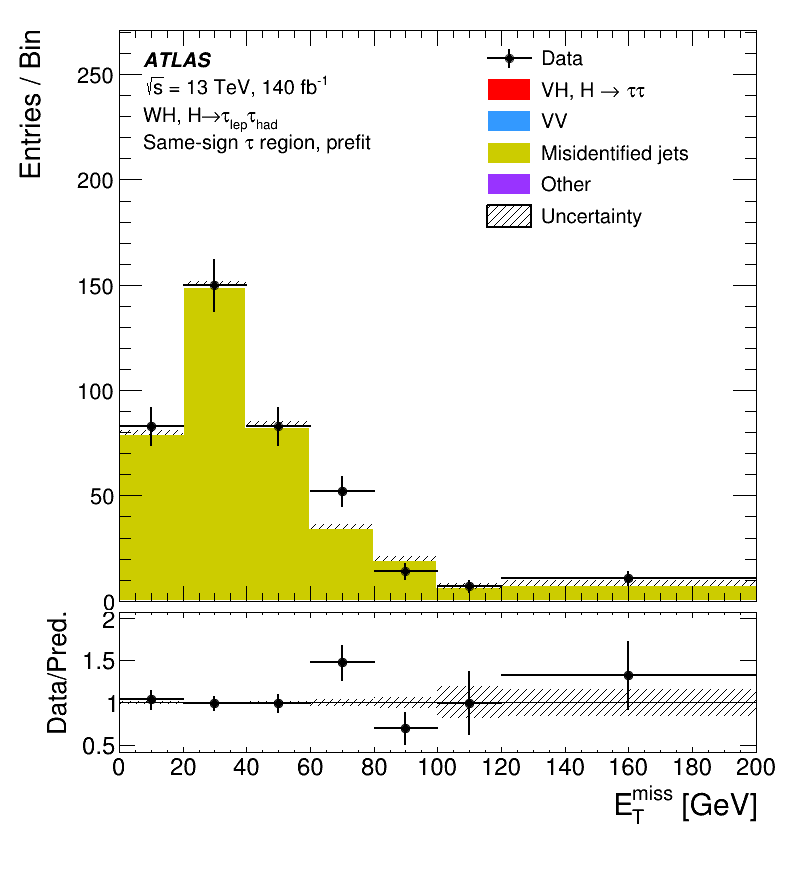} }\\
\subfloat[]{\includegraphics[width=0.43\textwidth]{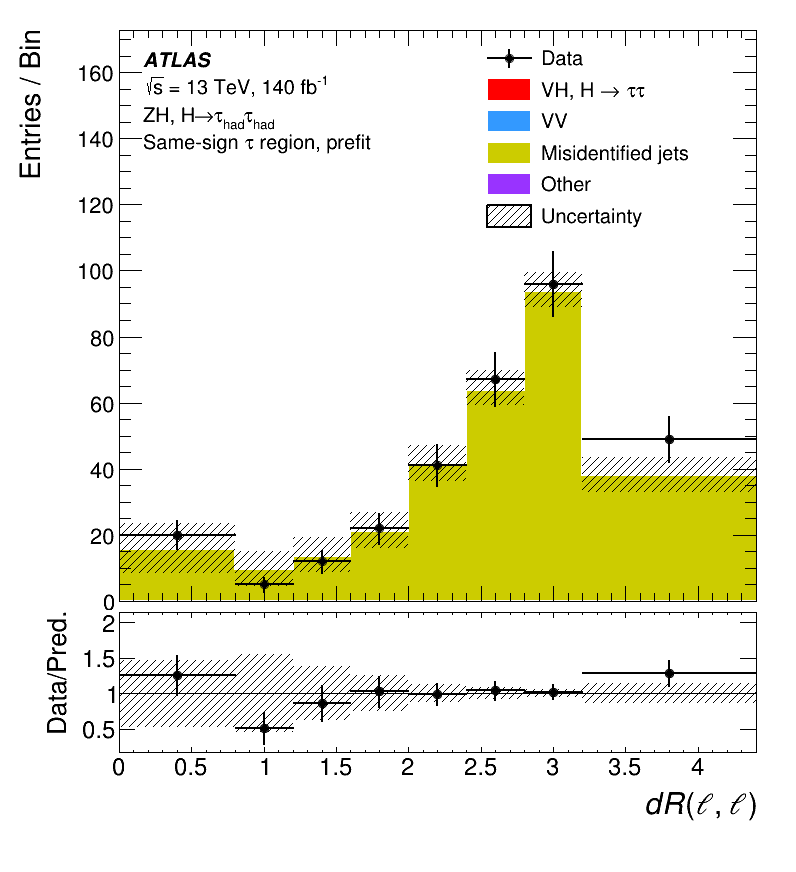}}
\subfloat[]{\includegraphics[width=0.43\textwidth]{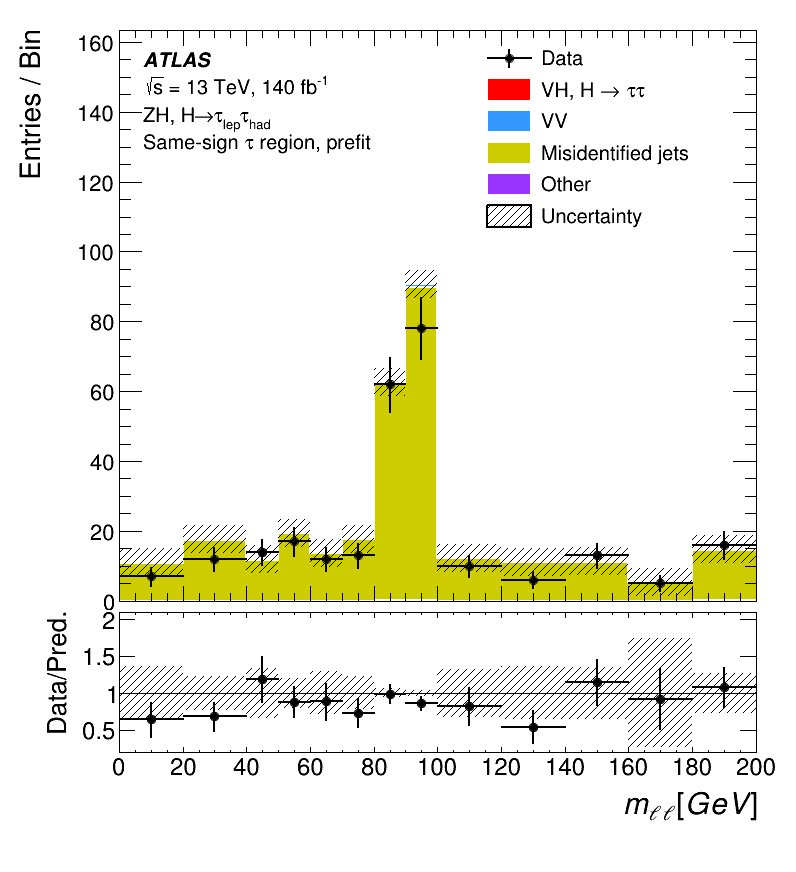}}
\caption{Distributions of representative kinematic variables in the
misidentified background-enriched same-sign region: (a) the Higgs
boson transverse momentum ($\pT^H$) in the \whhadhadshort{}
category, (b) the missing transverse momentum ($\met$) in the
\whlephadshort{} category, (c) the radial distance (dR$(\ell,\ell)$)
between the two light leptons associated to the \Zll{} decay process
in the \zhhadhadshort{} category, and (d) the invariant mass
($m_{\ell\ell}$) of the two light leptons associated to the \Zll{}
decay in the \zhlephadshort{} category.  The hatched band represents
the pre-fit statistical, experimental and theoretical
uncertainties. The signal contributions are considered as part of
the predictions and are normalized as predicted by the Standard
Model.}
\label{fig:ss_region}
\end{figure}

Figure~\ref{fig:ss_region} shows the distributions of some kinematic
variables that demonstrate the good modelling of the \fakebkg{} in the
misidentified background-enriched same-sign region in each of the four
analysis categories.



\section{Analysis strategy}
 
In this analysis, the signal strength is measured by a fit to a neural
network (NN) classifier score distribution. As a cross-check for this
method, an alternative strategy is applied using the same mass-based
analysis methods as in Run~1: \mmmc{} for the $ZH$ channels and
\mtwot{} for the $WH$ channels.

\subsection{Neural network analysis}\label{sec:nn_analysis}
The result is extracted using a fit to the distribution of the score
of a NN classifier. Six different NN classifiers are trained: one for
each of the $ZH$ and the \whhadhadshort{} channels and three for \whlephadshort{}, which has a
dedicated classifier for each combination of final light lepton
flavour ($e+e$, $\mu+\mu$, and $e+\mu$).  The NNs are trained to
distinguish simulated signal and diboson events using a combination of
low-level and high-level kinematic information from the particles in
the event (e.g. $p_T$, $|\eta|$, $\phi$) and the overall event
(e.g. \met{} and dilepton mass). The full list of the input variables
used for each NN is provided in Table~\ref{tab:ML_inputvar}.
The NNs training is performed at the \textsc{Preselection} stage to
overcome the limited statistics of the MC samples and take advantage
of the larger dataset size to avoid overtraining.
The mass-based observables \mmmc{} and \mtwot{} were not included in
the list of inputs, as it was determined that they did not yield a
significant improvement in sensitivity.
 
Each NN is implemented using \textsc{Keras}~\cite{keras} with a
\textsc{Tensorflow}~\cite{tensorflow} backend. The networks consist
of two initial transformation layers followed by three fully connected
layers of 128 nodes with \textsc{ReLU}~\cite{relu} activations.  The
output layer consists of a single node with a sigmoid activation
function. The two initial transformation layers enforce rotation
invariance in $\phi$ by adding a global $\phi$ offset during training,
which is consistently applied event--by--event to all
reconstructed objects.

\begin{table}[htb!]
\resizebox{1.\textwidth}{!}{
\begin{tabular}{c|c|c|c}
\textbf{All categories}  & \textbf{\zhhadhad}        & \textbf{\zhlephad}       & \textbf{\whhadhad}     \\ \hline
N-prongs($\tau_{1}$)     & N-prongs($\tau_{2}$)      & $\pT(\ell_2)$               & N-prongs($\tau_{2}$) \\
$\pT(\tau_{1})$           & $\pT(\tau_{2})$            & $\eta(\ell_2)$              & $\pT(\tau_{2})$             \\
$\eta(\tau_{1})$          & $\eta(\tau_{2})$           & $\phi(\ell_2)$             & $\eta(\tau_{2})$            \\
$\phi(\tau_{1})$          & $\phi(\tau_{2})$           & $\pT(H)$       & $\phi(\tau_{2})$            \\
$\Delta R (\tau_{1}, \ell_1)$& $\pT(\ell_2)$                 & $\eta(\ell_\tau)$           &$\sqrt{\eta(\ell_1)^2+\phi(\ell_1)^2}$\\
$\pT(l_1)$               & $\eta(\ell_2)$                & $\phi(\ell_\tau)$           &                            \\
$\eta(\ell_1)$              & $\phi(\ell_2)$                & $\Delta R (\ell,\ell)$        &                            \\
$\phi(\ell_1)$              & $m_{\ell \ell}$                    & $m_{\ell \ell}$                 &                            \\
$\pT(\met)$              & $\Delta R (\ell,\ell)$           &                         &                            \\
$\phi(\met)$             &                            &                         &                            \\  \cline{2-4}
& \multirow{2}{*}{\textbf{\whlephadee}}       & \multirow{2}{*}{\textbf{\whlephademu}}   & \multirow{2}{*}{\textbf{\whlephadmumu}}  \\
&                            &                         & \\\cline{2-4}
& $\pT(\ell_\tau)$               & $\pT(\ell_\tau)$            & $\pT(\ell_\tau)$  \\
& $\eta(\ell_\tau)$              & $\eta(\ell_\tau)$           & $\eta(\ell_\tau)$ \\
& $\phi(\ell_\tau)$              & $\phi(\ell_\tau)$           & $\phi(\ell_\tau)$ \\
& $\Delta \eta (\ell,\ell_\tau)$        & $\Delta \eta (\ell,\ell_\tau)$     & $\Delta \eta (\ell,\ell_\tau)$ \\
& jet width($\tau_{1}$)       & jet width($\tau_{1}$)    & jet width($\tau_{1}$) \\
& $\pT(H)$          & $m(\tau_1,\ell_\tau)$       & $\Delta R (\ell,\ell_\tau)$        \\
& $m(\tau_1,\ell_\tau)$          &  $\Delta R (\ell,\ell_\tau)$                & $m(\tau_1,l_\tau)$               \\
& $\Delta \eta(\tau_1,\ell_\tau)$& $\Delta \eta(\tau_1,\ell_\tau)$& $\Delta \eta(\tau_1,\ell_\tau)$      \\
& $\Delta \phi(l_1,\ell_\tau)$   & $\sum \pT(\mathrm{all\ visible})$ & $\Delta R(\tau_1,\ell_\tau)$         \\
& $\Delta_\phi(\tau_1,\ \met)$& $\Delta \phi(\tau_1,\met)$   &  $\sum \pT(\mathrm{all\ visible})$  \\
& $\Delta R (\ell,\ell_\tau)$           &         & $\Delta \phi(\ell_1,\ell_\tau)$      \\

\end{tabular}
}
\caption{Input variables for the neural networks included in all
channels, and then for the specific category. The indexes ``1'' and
``2'' refer to the leading and sub-leading objects, respectively
(following a $\pT$ ordering). The symbol
$\ell_\tau$ refers to the light lepton originating from a
$\tau$-lepton decay, while $\ell$ (without any index) refers to a
light lepton associated with the $V$ boson
decay. \label{tab:ML_inputvar}}
\end{table}
 
The binning of the NN score distributions in the four
categories results from an optimization process that maximized significance under the constraint that the statistical uncertainty associated with the signal and
background templates is no larger than 20\% in each bin.

\subsection{Mass-based analysis}\label{sec:mass_analysis}
As a cross-check for the main analysis strategy, a historical approach used in the Run 1 analysis based on the mass observables
(\mmmc{} for the $ZH$ channels and the \mtwot{} for the $WH$
channels) is also adopted to extract the signal yield. The $WH$ categories use \mtwot{} instead of
\mmmc{} because the presence of
an additional neutrino coming from the $W$ decay breaks an important
assumption used by the \mmmc{} algorithm, as previously discussed in
Section~\ref{sec:evt_cat}.

For the mass-based analysis, the cut on the mass observable
is dropped from the \textsc{Signal} selection criteria in all categories to
better constrain the background using the sidebands in which the
background contribution is dominant. As a consequence, the total
number of events selected in the \textsc{Signal} region increases up to a
factor of about four due to a general increase of the background
fraction from \fake.
 
As shown in Table~\ref{tab:ML_inputvar} , the mass variables
defined here (\mmmc{} and \mtwot) are not used as inputs for the NN.
However, the binning of the distributions of the mass variables follows the same
optimization criteria used in the NN-based fit case discussed in the
previous section.


\section{Systematic uncertainties}
\label{sec:syst}
 
Systematic uncertainties affect the yields in the signal and control
regions as well as the shape of the fitted distribution.  They can be
separated into four groups: MC sample statistical uncertainties (using
the lite version of the Beeston-Barlow method~\cite{Cranmer:2012sba}),
experimental uncertainties, theoretical uncertainties for the
backgrounds, and theoretical uncertainties for the signal.  The
systematic uncertainties related to the estimation of misidentified
objects are described in Section~\ref{sec:background}.
 
Experimental uncertainties pertain to the trigger as well as
final-state objects: reconstruction, identification and isolation
efficiency uncertainties for electrons~\cite{EGAM-2021-01},
muons~\cite{PERF-2015-10}, {\tauhadvis}~\cite{ATLAS-CONF-2017-029},
jets~\cite{PERF-2016-04,ATLAS-CONF-2014-018,PERF-2016-06,JETM-2018-05},
\emph{b}-jets~\cite{PERF-2016-05,FTAG-2019-02} and
\met~\cite{PERF-2016-07}.
The uncertainties associated with the \tauhadvis{} identification
efficiency are in the range of 2\% to 6\%, while the \textsc{eBDT}
efficiency uncertainty is 1\% to 2\%. All these uncertainties are
parameterised as a function of the \tauhadvis{} \pt{} and number of
associated tracks or $\tau$--lepton decay mode (\textsc{eBDT}
efficiency). For the \tauhadvis{} energy scale, the total uncertainty
is in the range of 1\% to 4\%, arising from a combination of
measurements: (i) a direct measurement with $Z\to\tau\tau\to
\mu\tauhadvis + 3 \nu$ events, (ii) measurements of the calorimeter
response to single particles, and (iii) comparisons between
simulations using different detector geometries or \GEANT physics
lists~\cite{Agostinelli:2002hh}. This uncertainty is also
parameterised as a function of the \tauhadvis{} \pt{} and number of
associated tracks~\cite{ATLAS-CONF-2017-029}.
The energy scale and resolution uncertainties of final state objects
are taken into account as well.  Experimental uncertainties affect the
shape of the distribution of the final discriminant, the background
yields, and the signal cross-section through their effects on the
acceptance of and migration between different analysis categories.
An additional uncertainty from the measurement of the
luminosity~\cite{DAPR-2021-01,LUCID2}, amounting to 0.83\%, is also
included.
 
The theoretical uncertainties for the diboson, triboson and the
top quark background are estimated from simulation. These include the
systematic uncertainties due to renormalisation (\muR), factorisation
(\muF) and resummation scale (\qsf), the jet-to-parton matching scheme
(\ckkw)~\cite{CKKW}, the choice of \alphas value, and the PDFs.
For the diboson background, the most relevant contributions come from
the systematic uncertainties due to \muR{} and \muF, which affect the
shape and the global normalization with a total uncertainty that
ranges from 7\% to 12\% in the in the $ZZ$ and $WZ$ processes
respectively.
For the top quark background, uncertainties related to the choice of
matrix element and parton shower
generators~\cite{Bahr:2008pv,Bellm:2015jjp}, the initial- and
final-state radiation model~\cite{ATL-PHYS-PUB-2017-007}, and the PDFs
are considered~\cite{Harland-Lang:2014zoa}. Their effect on the
normalisation and shape of the final discriminant is considered in the
statistical analysis.
 
The Higgs boson production cross-section uncertainties are obtained
from Ref.~\cite{deFlorian:2016spz}.  
To account for missing higher orders in QCD, additional uncertainties
are estimated by varying \muR, \muF, \qsf, the choice of \alphas
value, and the choice of matrix element generator or parton shower and
hadronisation model. For the matrix element variation, predictions by
\POWHEGBOX[v2] are compared with those by
\MGNLO~\cite{Alwall:2014hca}. The parton shower and hadronisation
model variation replaces the nominal \PYTHIA[8] simulation with
\HERWIG[7]~\cite{Bellm:2015jjp,Bahr:2008pv}. Additional theoretical
uncertainties affecting the \ttH production cross-section are also
considered; more details can be found in Ref.~\cite{ATLAS:2022yrq}.


\section{Results}
The statistical procedure is based on a likelihood function
$\mathcal{L}(\mu,\theta)$, constructed as the product of Poisson
probability terms over the bins of the input distributions. The
parameter of interest, $\mu$, is the signal strength that multiplies
the SM Higgs boson production cross-section in association with a
vector boson times the branching fraction into $\tau\tau$. It is
extracted by maximising the likelihood. An additional statistical
procedure, also based on a likelihood function defined as described
above, is used to estimate two parameters of interest $\mu_{ZH}$ and
$\mu_{WH}$ separately for the $ZH$ and the $WH$ categories,
respectively.
Systematic uncertainties enter the likelihood as nuisance parameters
(NP), $\theta$. Most of the uncertainties discussed in
Section~\ref{sec:syst} are constrained with Gaussian or log-normal
probability density functions.
The systematic variations that are subject to large statistical
fluctuations are smoothed, and systematic uncertainties that have a
negligible impact on the final results are pruned away
category by category. 
Only the \textsc{Signal} regions are considered in the fit. The normalisation
of the background contributions from diboson, triboson, top processes
and other small backgrounds are taken from the simulation.
The probability that the background-only hypothesis is compatible with
the observed data is determined using the profile-likelihood ratio
test statistic defined in Ref.~\cite{Cowan:2010js}.
 
\subsection{Results of the neural network analysis}\label{sec:nn_results}
For a Higgs boson mass of 125~GeV, when the $ZH$ and $WH$ categories
are combined under the constraint that
$\mu_{VH}^{\tau\tau}=\mu_{ZH}^{\tau\tau}=\mu_{WH}^{\tau\tau}$, the
NN-based fit shows an observed significance of 4.2 standard deviations
from the background-only hypothesis, compared with an expectation of
3.6 standard deviations. The fitted value of the signal strength is:
\begin{equation}
\mu_{VH}^{\tau\tau} = 1.28\ ^{+0.39}_{-0.36} = 1.28\ ^{+0.30}_{-0.29}\ (\mathrm{stat.})\ ^{+0.25}_{-0.21}\ (\mathrm{syst.})  \nonumber.
\end{equation}
 
Using a predicted cross-section of $(6.59 \pm 0.03)$~fb from the
Standard Model~\cite{deFlorian:2016spz}, this corresponds to a
measured cross-section of $8.5^{+2.6}_{-2.4}$~fb.
 
The total statistical uncertainty is defined as the uncertainty in
$\mu_{\text{VH}}^{\tau\tau}$ when all the NPs are fixed to their
best-fit values. The total systematic uncertainty is then defined as
the difference in quadrature between the total uncertainty in
$\mu_{\text{VH}}^{\tau\tau}$ and the total statistical
uncertainty. The result obtained is limited by the data sample size,
as shown by the breakdown of the statistical and systematic
uncertainties.
 
The relative effects of systematic uncertainties on the measurement of
$\mu_{\text{VH}}^{\tau\tau}$ are shown in Table~\ref{tab:impact_1POI}. The
impact of a category of systematic uncertainties is defined as the
difference in quadrature between the uncertainty in
$\mu_{\text{VH}}^{\tau\tau}$ computed when all NPs are fitted and that when
the NPs in the category are fixed to their best-fit values.
As shown in Table~\ref{tab:impact_1POI}, the systematic uncertainties
associated to the \tauhadvis{} reconstruction (including its
identification and calibration) and the systematic uncertainties
associated to the background sample size play a dominant role,
followed by the modelling of the \fakebkg{}.
\begin{table}[htb!]
\centering
\caption{Summary of the different sources of uncertainty affecting the
observed $\mu_{\text{VH}}^{\tau\tau}$ and their impact as computed
by the NN-based fit described in Section~\ref{sec:nn_analysis}.
Experimental uncertainties for reconstructed objects combine
efficiency and energy/momentum scale and resolution
uncertainties. \enquote{Simulated background sample size} includes
the bin-by-bin statistical uncertainties in the simulated
backgrounds and in misidentified jets background, which is estimated
using data.  }
\label{tab:impact_1POI}
\begin{tabular}{l|c}
\toprule
Source of uncertainty      & $\delta \mu/\mu_{\text{VH}}^{\tau\tau}$ [\%] \\ 
\hline
Hadronic $\tau$--lepton decay       &  9  \\
Simulated background sample size    &  9     \\
Misidentified jets         &  4   \\
Jet and \met               &  4  \\
Theoretical uncertainty in signal                                &  4  \\
Theoretical uncertainty in top-quark, $VV$ and $VVV$ processes   &  4  \\
Electrons and muons        &  2  \\
Luminosity                 &  1  \\
Flavour tagging            &  $<1$ \\
\hline
Total systematic uncertainty       &  16     \\
Total statistical uncertainty      &  24 \\
\hline
Total                              &  30 \\
\bottomrule
\end{tabular}
 
\end{table}
 
\begin{figure}[!htb]
\centering
\subfloat[]{\includegraphics[width=0.43\textwidth]{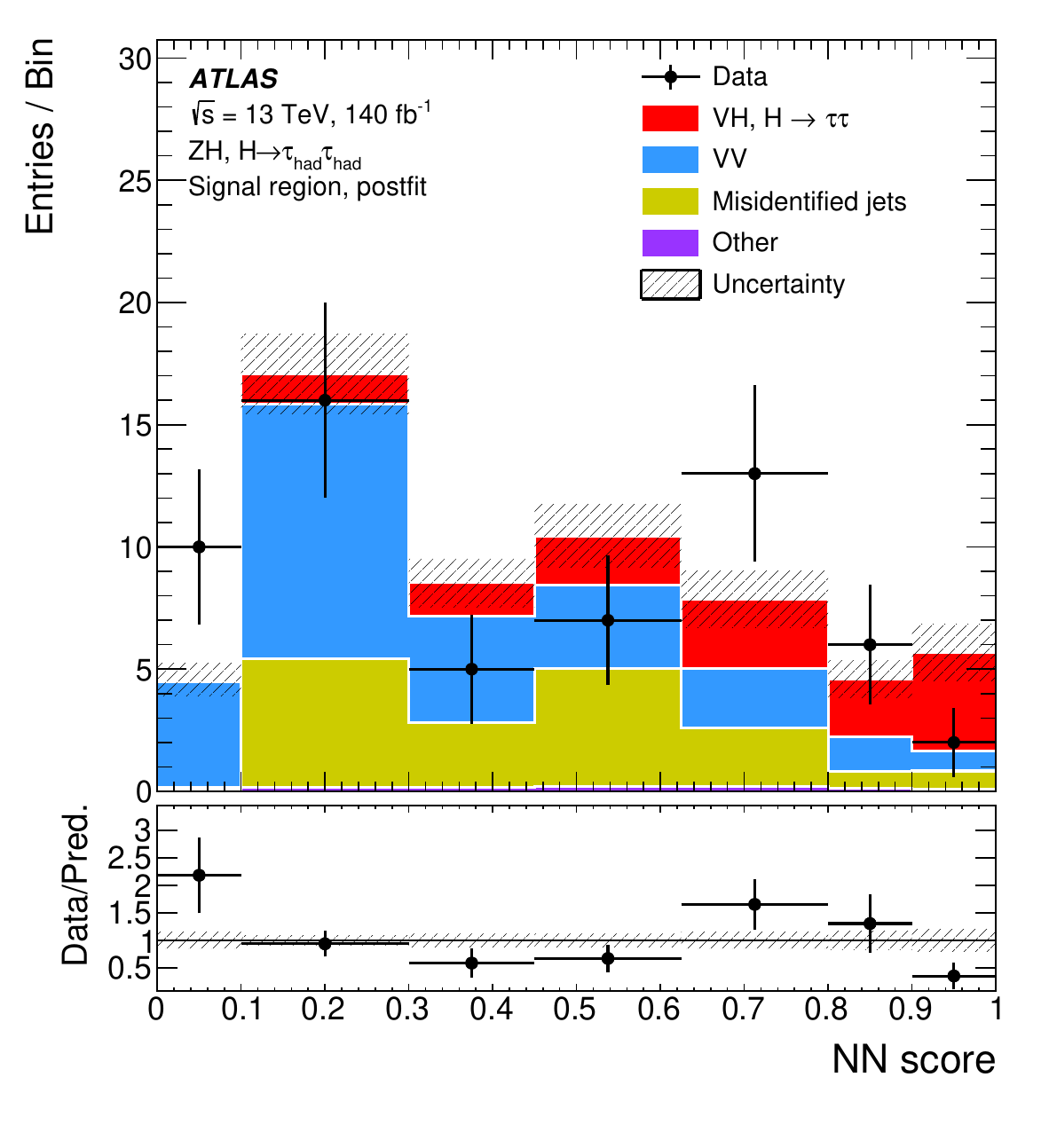}}
\subfloat[]{\includegraphics[width=0.43\textwidth]{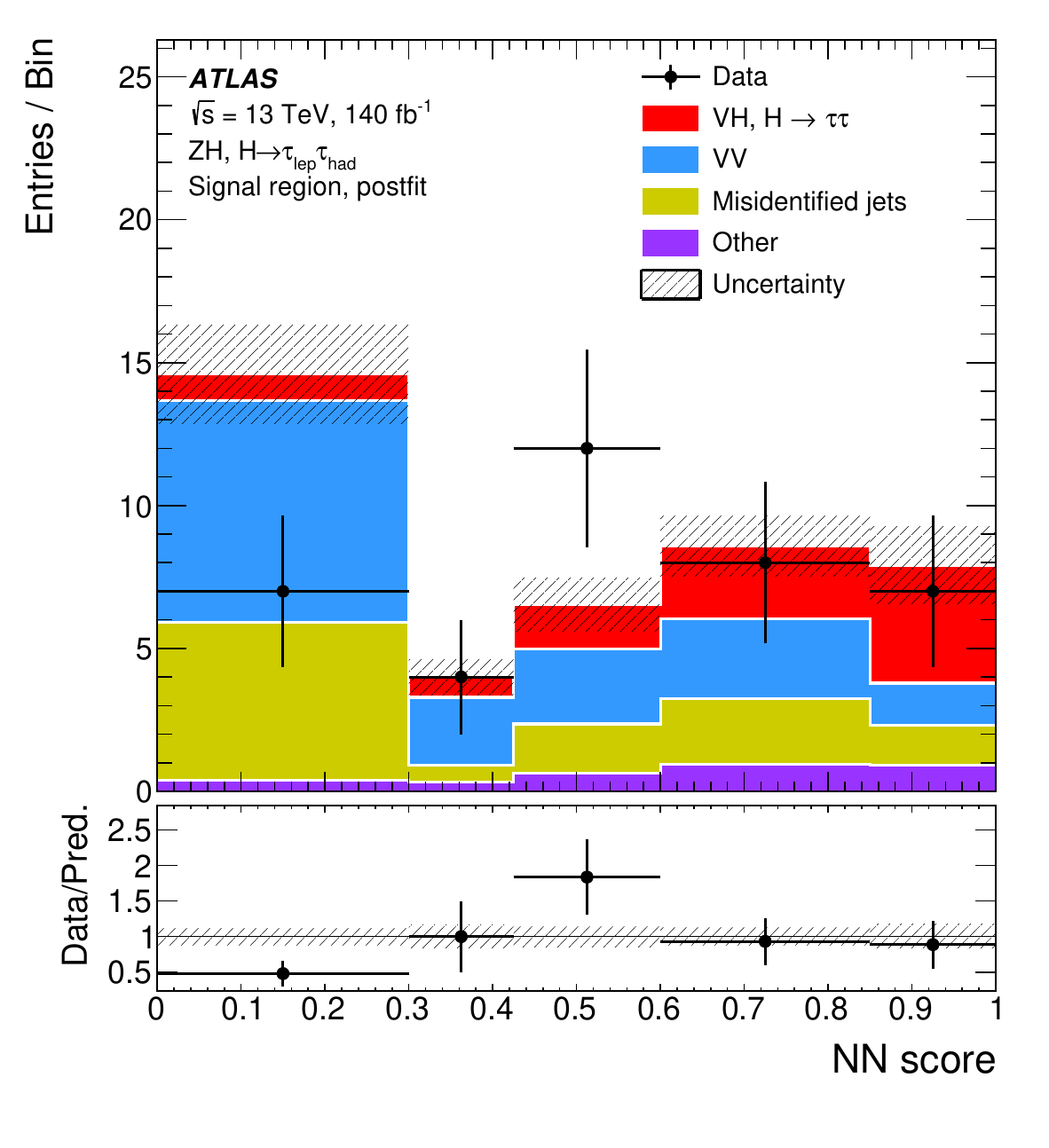}} \\
\subfloat[]{\includegraphics[width=0.43\textwidth]{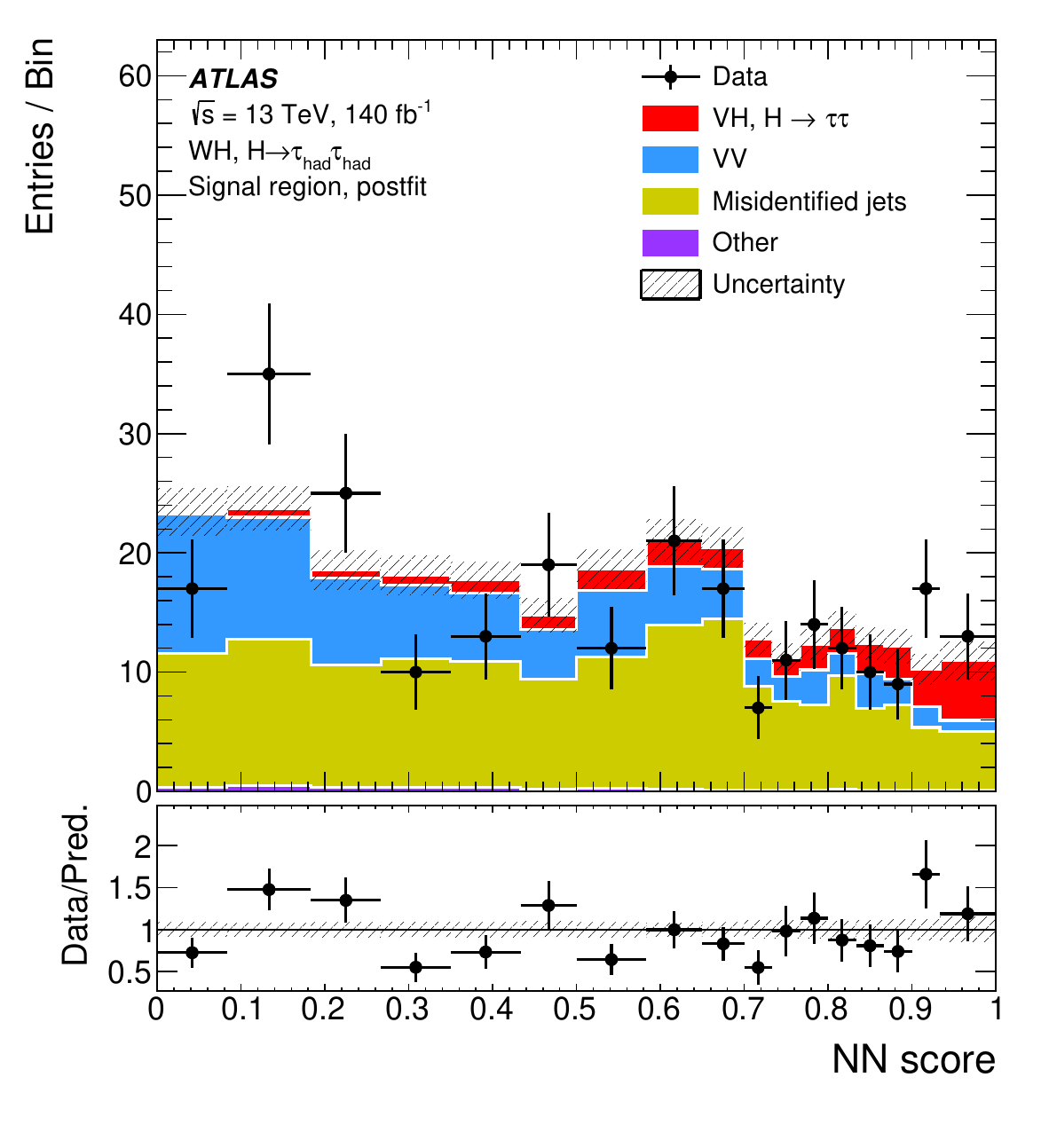}}
\subfloat[]{\includegraphics[width=0.43\textwidth]{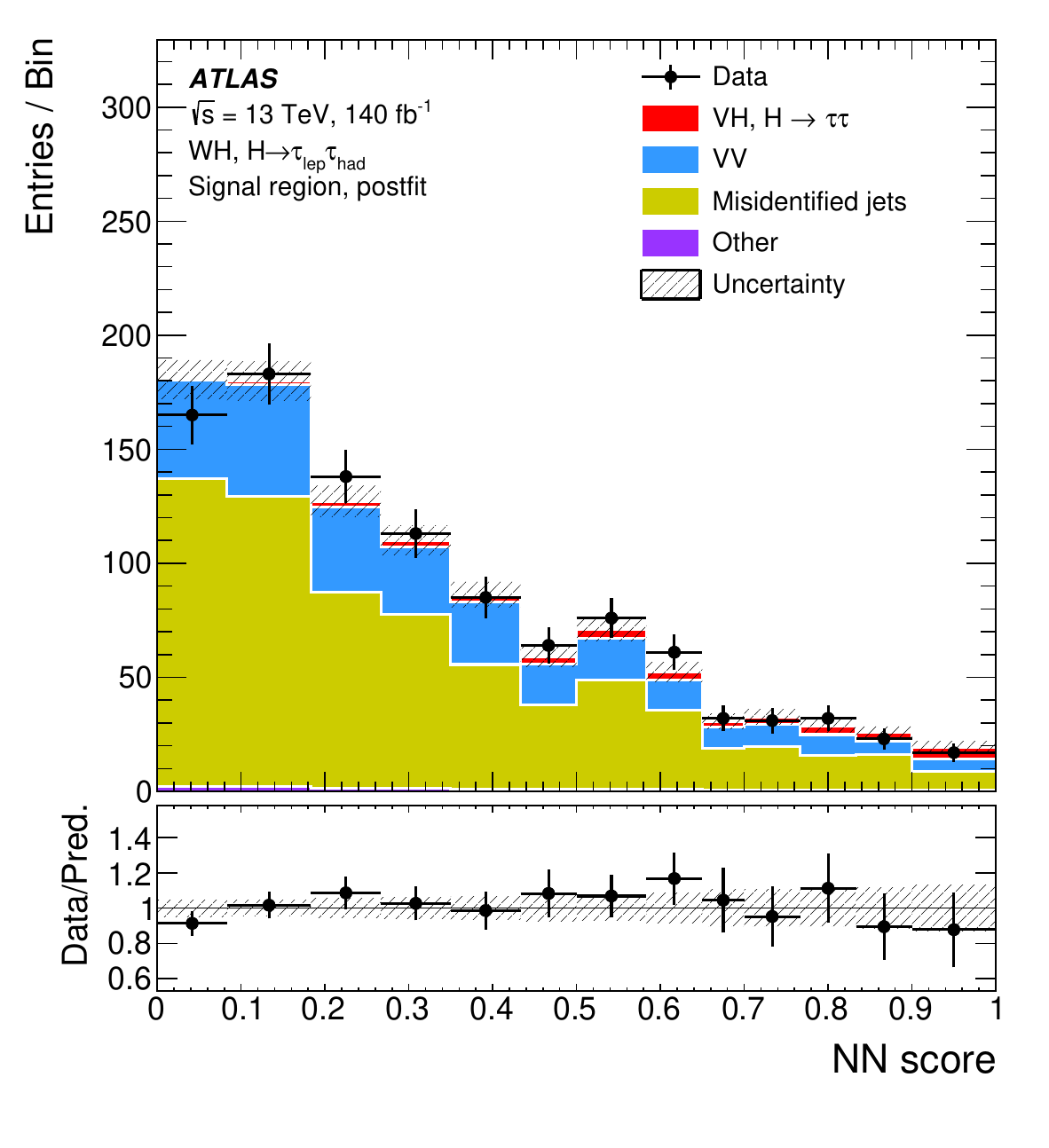}}
\caption{Post-fit distributions for NN-based analysis of the NN-scores
in the \zhhadhadshort{} (a), \zhlephadshort{} (b), \whhadhadshort{}
(c) and \whlephadshort{} (d) categories. The hatched band indicates
the total post-fit uncertainty of the total predicted yields. The
post-fit signal contributions are considered as part of the
predictions.}
\label{fig:nn_fit}
\end{figure}

Figure~\ref{fig:nn_fit} shows the post-fit distributions of the NN
scores. The background prediction in all post-fit distributions is
obtained by setting the nuisance parameters according to their
best--fit values. Significant studies into the goodness of fit were
performed given the fluctuations seen in several distributions,
including working with Monte Carlo simulation toys to confirm the
applicability of the asymptotic approximation used in extracting the
analysis sensitivity.
 
\begin{figure}[!htb]
\centering
\includegraphics[width=0.43\textwidth]{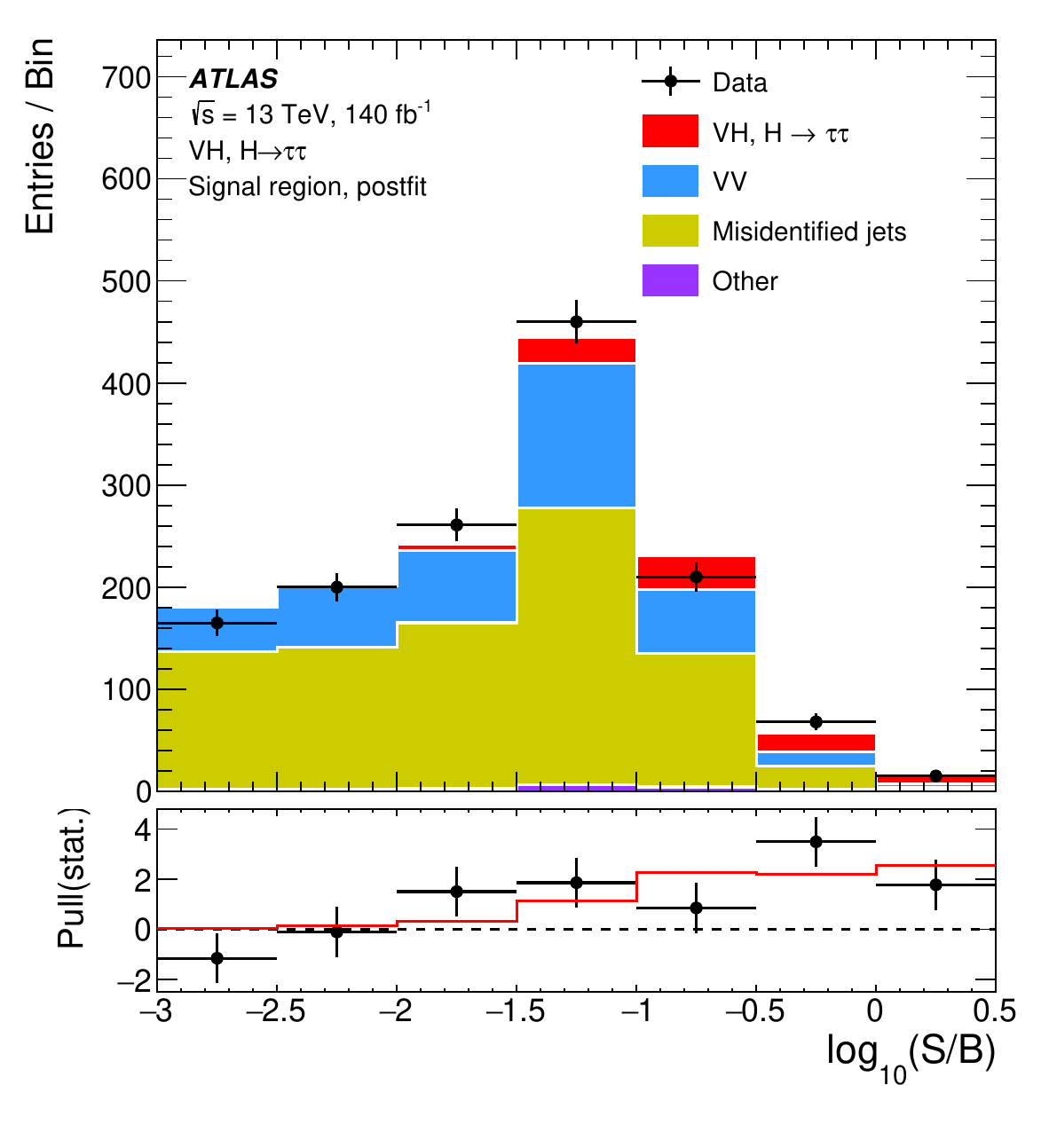}
\caption{ Event yields as a function of $\log_{10}(S/B)$ for data,
background, and a Higgs boson signal with $m_{H} =
125$~GeV. Final-discriminant bins in all regions are combined into
bins of $\log_{10}(S/B)$, where $S$ is the fitted signal and $B$ is
the fitted background from the NN-based fit. The Higgs boson signal
contribution is shown after re-scaling the SM cross-section
according to the value of the signal strength extracted from data
($\mu = 1.28$). In the lower panel, the pull of the data relative to
the background-only expectation, evaluated as the difference between
the data and the background-only expectation divided by the
square-root sum of the data and background statistical uncertainty,
is shown. The solid line shows the expected pull in each bin for
the best-fit signal value.
}
\label{fig:nn_fit_sOveB}
\end{figure}

Figure~\ref{fig:nn_fit_sOveB} shows the data, background, and signal
yields, where final-discriminant bins in all regions are combined into
bins of $\log_{10}(S/B)$. Here, $S$ and $B$ are the fitted signal and
background yields in each analysis bin.
Table~\ref{tab:nn_yields} shows the good agreement obtained in the
resulting yields in the four categories.
\begin{table}[htb!]
\resizebox{.75\textwidth}{!}{
\begin{tabular}{l|r @{${}\pm{}$} l|r @{${}\pm{}$} l|r @{${}\pm{}$} l|r @{${}\pm{}$} l}
\hline
& \multicolumn{2}{c|}{\zhhadhadshort} & \multicolumn{2}{c|}{\zhlephadshort} & \multicolumn{2}{c|}{\whhadhadshort} & \multicolumn{2}{c}{\whlephadshort}\\
\hline\hline
$ZH$          & 14 & 4 & 10 & 3 & \multicolumn{2}{c|}{-} & \multicolumn{2}{c}{-} \\
$WH$          & \multicolumn{2}{c|}{-} & \multicolumn{2}{c|}{-} & 31 & 9 & 40 & 12 \\
Misidentified jets      & 17 & 2 & 12 & 1 & 160 & 7 & 677 & 31 \\
top-quark     & 0.35 & 0.06 & 2.3 & 0.4 & 0.64 & 0.11 & 3.2 & 1.6 \\
$VVV$         & 0.09 & 0.02 & 0.43 & 0.07 & 0.46 & 0.07 & 6 & 1 \\
$ZZ$          & 27 & 3 & 17 & 2 & \multicolumn{2}{c|}{-} & 0.06 & 0.04 \\
\ttH          & 0.51 & 0.07 & 0.44 & 0.07 & 2.49 & 0.77 & 3.86 & 1.19 \\
$WZ$          & \multicolumn{2}{c|}{-} & \multicolumn{2}{c|}{-} & 76 & 10 & 273 & 26 \\
\hline
Total         & 59 & 5 & 42 & 4 & 272 & 12 & 1003 & 28 \\
\hline
Data          & \multicolumn{2}{c|}{59} & \multicolumn{2}{c|}{38} & \multicolumn{2}{c|}{262} & \multicolumn{2}{c}{1020} \\
\hline
\end{tabular}
}
\caption{Post-fit yields from the NN-based fit performed with
$\mu_{VH}^{\tau\tau}=\mu_{ZH}^{\tau\tau}=\mu_{WH}^{\tau\tau}$. The
symbol ``-'' is used when no events or $<10^{-2}$ events are present.}
\label{tab:nn_yields}
\end{table}
 
\begin{figure}[!htb]
\centering
\includegraphics[width=0.653\textwidth]{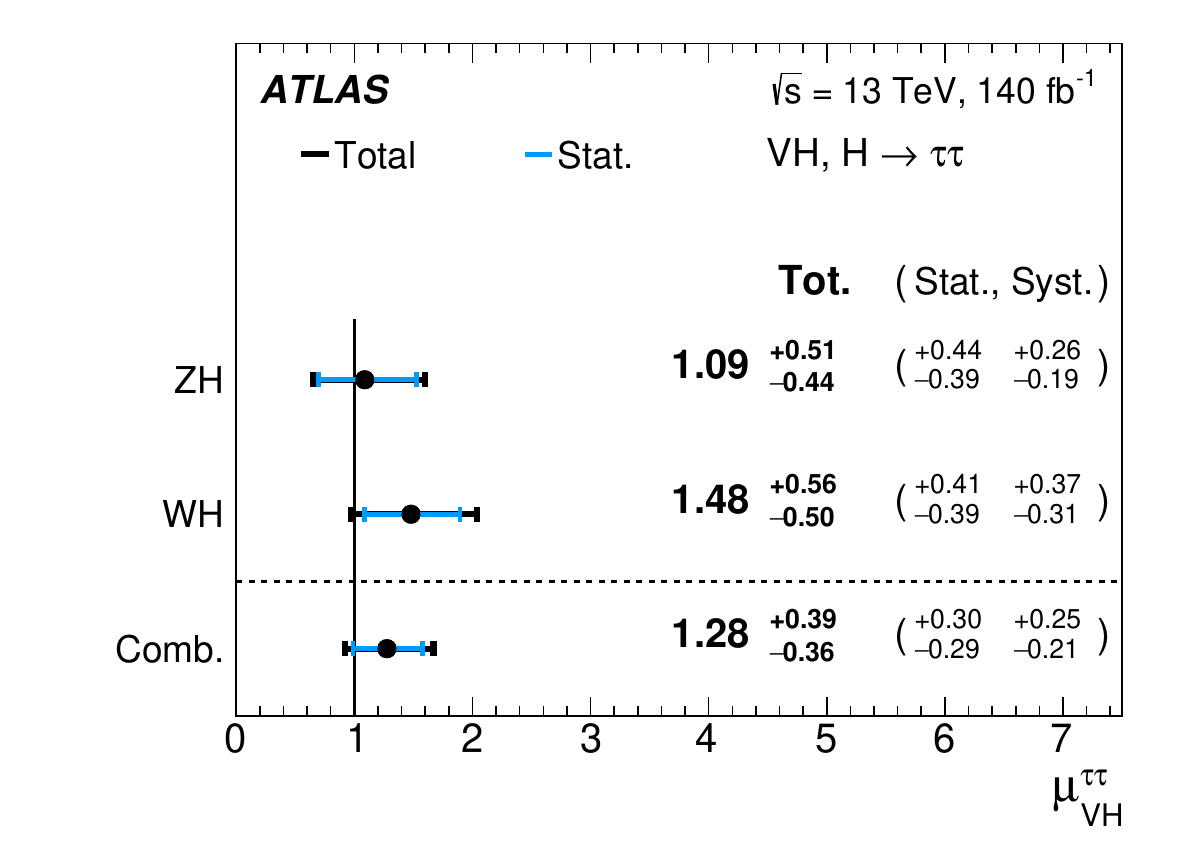}
\caption{ The fitted values of the Higgs boson signal strength
$\mu_{VH}^{\tau\tau}$ for $m_{H} = 125$~GeV for the $WH$ and $ZH$
processes and their combination from the NN-based fit. The
individual $\mu_{VH}^{\tau\tau}$ values for the $(W/Z)H$ processes
are obtained from a simultaneous fit with the signal strength for
each of the $WH$ and $ZH$ processes floating independently. The
probability of compatibility of the individual signal strengths is
56\%.}
\label{fig:nn_fit_mu}
\end{figure}

A combined fit is also performed with floating signal strengths
separately for the $WH$ and $ZH$ production processes. The results of
this fit are shown in Figure~\ref{fig:nn_fit_mu}. The probability that
the signal strengths measured in the two production processes are
compatible is 56\%\footnote{The compatibility between fits differing
only in their number of parameters of interest is evaluated in the
asymptotic regime, where the difference between their maximum
likelihoods follows a $\chi^2$ distribution with a number of degrees
of freedom equal to the difference between the numbers of parameters
of interest.}.
 
\FloatBarrier
 
\subsection{Results of the mass-based analysis}
For all channels combined, the fitted value of the signal strength is:
\begin{equation}
\mu_{VH}^{\tau\tau} = 1.40\ ^{+0.49}_{-0.45} = 1.40\  ^{+0.36}_{-0.35}\ (\mathrm{stat.})\ ^{+0.33}_{-0.28}\ (\mathrm{syst.})  \nonumber
\end{equation}
in good agreement with the result of the NN-based analysis discussed
in the previous section. The observed excess in the mass analysis has
a significance of 3.5 standard deviations, compared to an expectation
of 2.6 standard deviations. 
The relative effects of systematic uncertainties on the measurement of
$\mu_{\text{VH}}^{\tau\tau}$ in this case are quite similar to the
results discussed in Section~\ref{sec:nn_results}.
 
\begin{figure}[!htb]
\centering
\subfloat[]{\includegraphics[width=0.43\textwidth]{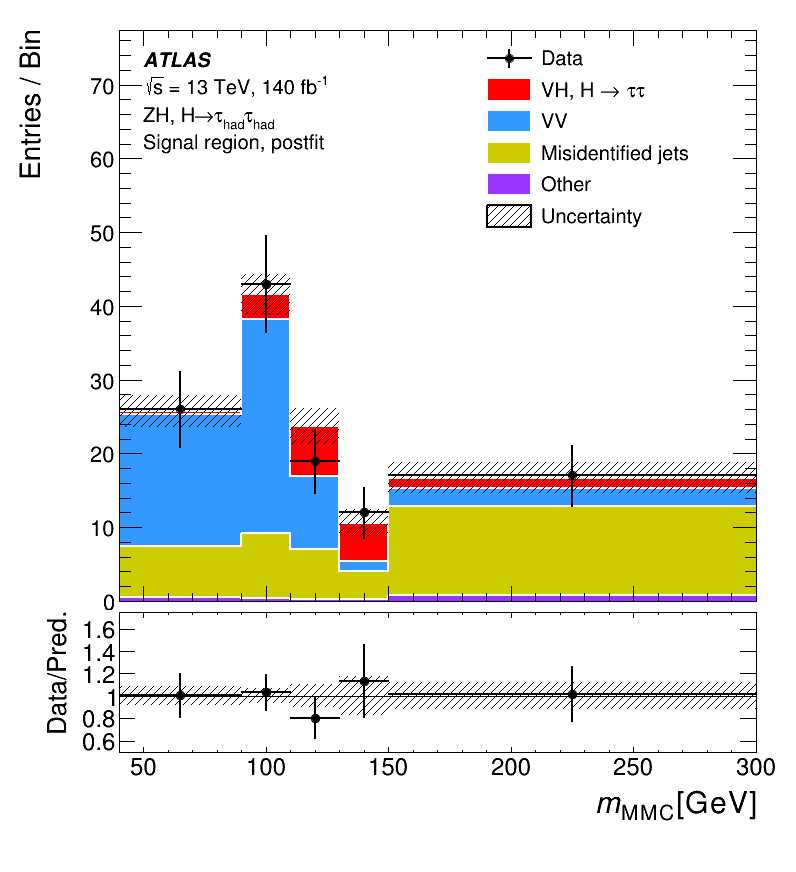}}
\subfloat[]{\includegraphics[width=0.43\textwidth]{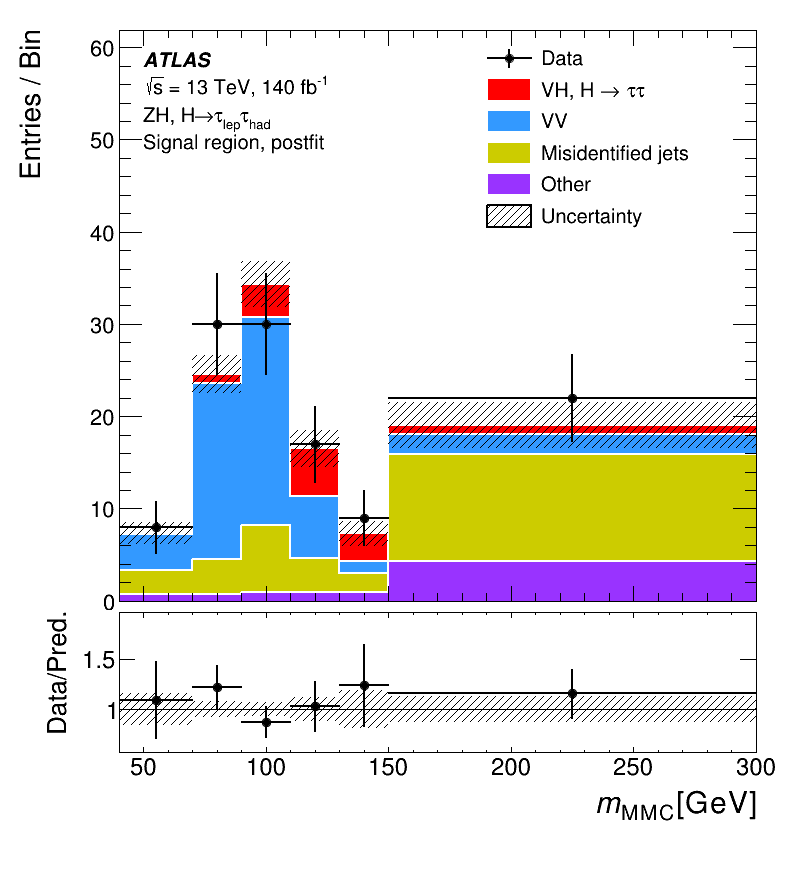}} \\
\subfloat[]{\includegraphics[width=0.43\textwidth]{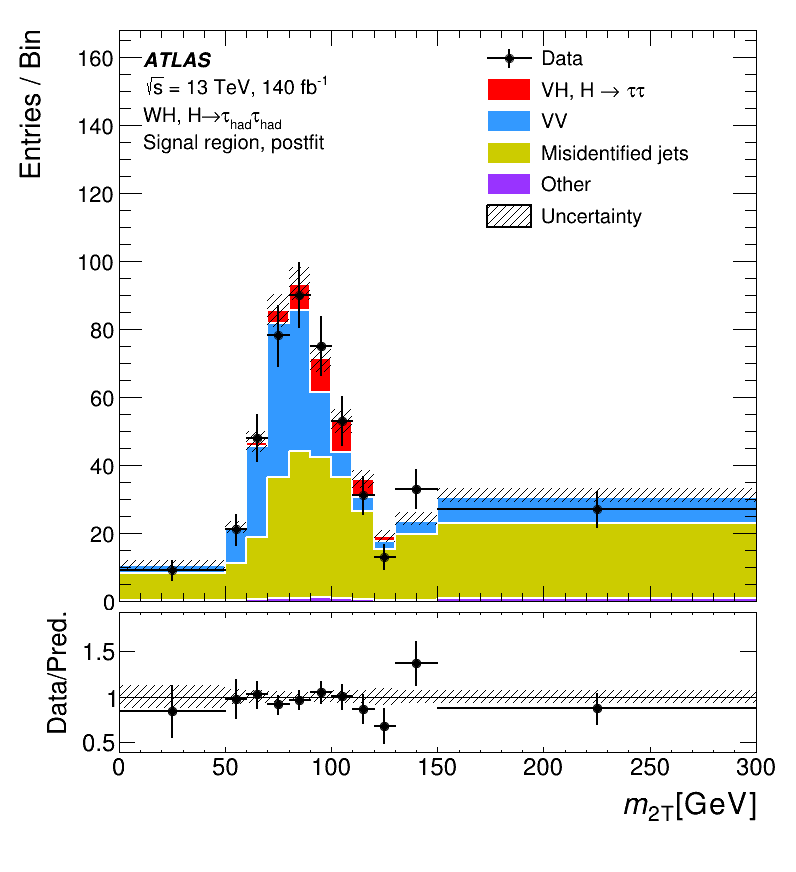}}
\subfloat[]{\includegraphics[width=0.43\textwidth]{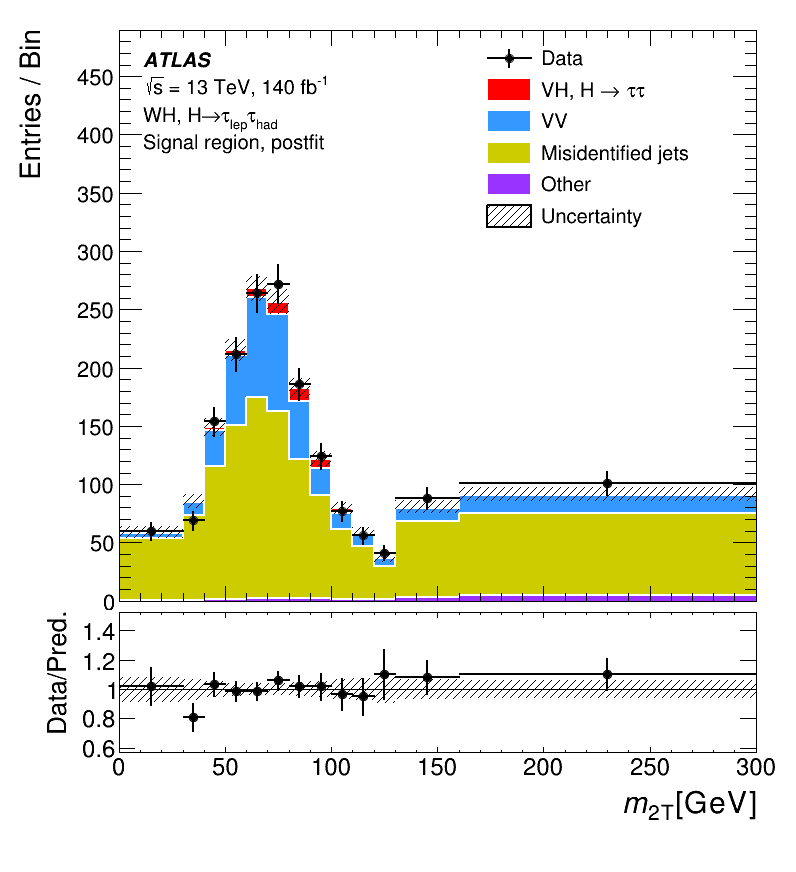}}
\caption{Post-fit distributions for mass-based analysis for \mmmc{}
in the (a) \zhhadhadshort, (b) \zhlephadshort, and for \mtwot{}
in the (c) \whhadhadshort{} and (d) \whlephadshort{} categories. The
hatched band indicates the total post-fit uncertainty of the total
predicted yields. The post-fit signal contributions are considered
as part of the predictions.}
\label{fig:mass_fit}
\end{figure}

Figure~\ref{fig:mass_fit} shows the post-fit distributions of the
observables used in the mass-based analysis: \mmmc{} for the $ZH$
categories and \mtwot{} for the $WH$ ones. The background prediction
in all post-fit distributions is obtained by setting the nuisance
parameters according to the results of the combined $ZH$ and $WH$ fit.


\FloatBarrier
 
\section{Conclusion}
 
A search for the Standard Model Higgs boson decaying into a $\tau\tau$
pair and produced in association with a leptonically-decaying $W$ or $Z$ boson is presented,
using data collected by the ATLAS experiment in proton--proton
collisions from Run 2 of the LHC. The data correspond to an integrated
luminosity of 140 fb$^{-1}$ collected at a centre-of-mass energy of
$\sqrt{s} =13$~TeV.
 
In addition to the approximately seven times larger dataset, the main
sources of improvement with respect to the Run 1 result are the more
sophisticated analysis methodologies. These include the
introduction of a neural network discriminator for rejecting the
diboson background and better $\tauhadvis$ identification algorithms.
 
An excess over the expected background is observed with a significance
of 4.2 standard deviations compared with an expectation of 3.6. The
measured signal strength relative to the SM prediction for $m_H =
125$~GeV is found to be $\mu_{\text{VH}} = 1.28\ ^{+0.30}_{-0.29}\
(\mathrm{stat.})\ ^{+0.25}_{-0.21}\ (\mathrm{syst.})$.
This analysis provides currently the most sensitive measurement of the
Higgs boson and a leptonically-decaying vector boson in events where
the Higgs boson decays into a pair of $\tau$-leptons.
 
\enlargethispage{\baselineskip}


\clearpage
\section*{Acknowledgements}


We thank CERN for the very successful operation of the LHC and its injectors, as well as the support staff at
CERN and at our institutions worldwide without whom ATLAS could not be operated efficiently.
 
The crucial computing support from all WLCG partners is acknowledged gratefully, in particular from CERN, the ATLAS Tier-1 facilities at TRIUMF/SFU (Canada), NDGF (Denmark, Norway, Sweden), CC-IN2P3 (France), KIT/GridKA (Germany), INFN-CNAF (Italy), NL-T1 (Netherlands), PIC (Spain), RAL (UK) and BNL (USA), the Tier-2 facilities worldwide and large non-WLCG resource providers. Major contributors of computing resources are listed in Ref.~\cite{ATL-SOFT-PUB-2023-001}.
 
We gratefully acknowledge the support of ANPCyT, Argentina; YerPhI, Armenia; ARC, Australia; BMWFW and FWF, Austria; ANAS, Azerbaijan; CNPq and FAPESP, Brazil; NSERC, NRC and CFI, Canada; CERN; ANID, Chile; CAS, MOST and NSFC, China; Minciencias, Colombia; MEYS CR, Czech Republic; DNRF and DNSRC, Denmark; IN2P3-CNRS and CEA-DRF/IRFU, France; SRNSFG, Georgia; BMBF, HGF and MPG, Germany; GSRI, Greece; RGC and Hong Kong SAR, China; ISF and Benoziyo Center, Israel; INFN, Italy; MEXT and JSPS, Japan; CNRST, Morocco; NWO, Netherlands; RCN, Norway; MEiN, Poland; FCT, Portugal; MNE/IFA, Romania; MESTD, Serbia; MSSR, Slovakia; ARRS and MIZ\v{S}, Slovenia; DSI/NRF, South Africa; MICINN, Spain; SRC and Wallenberg Foundation, Sweden; SERI, SNSF and Cantons of Bern and Geneva, Switzerland; MOST, Taipei; TENMAK, T\"urkiye; STFC, United Kingdom; DOE and NSF, United States of America.
 
Individual groups and members have received support from BCKDF, CANARIE, CRC and DRAC, Canada; CERN-CZ, PRIMUS 21/SCI/017 and UNCE SCI/013, Czech Republic; COST, ERC, ERDF, Horizon 2020, ICSC-NextGenerationEU and Marie Sk{\l}odowska-Curie Actions, European Union; Investissements d'Avenir Labex, Investissements d'Avenir Idex and ANR, France; DFG and AvH Foundation, Germany; Herakleitos, Thales and Aristeia programmes co-financed by EU-ESF and the Greek NSRF, Greece; BSF-NSF and MINERVA, Israel; Norwegian Financial Mechanism 2014-2021, Norway; NCN and NAWA, Poland; La Caixa Banking Foundation, CERCA Programme Generalitat de Catalunya and PROMETEO and GenT Programmes Generalitat Valenciana, Spain; G\"{o}ran Gustafssons Stiftelse, Sweden; The Royal Society and Leverhulme Trust, United Kingdom.
 
In addition, individual members wish to acknowledge support from CERN: European Organization for Nuclear Research (CERN PJAS); Chile: Agencia Nacional de Investigaci\'on y Desarrollo (FONDECYT 1190886, FONDECYT 1210400, FONDECYT 1230812, FONDECYT 1230987); China: National Natural Science Foundation of China (NSFC - 12175119, NSFC 12275265, NSFC-12075060); Czech Republic: PRIMUS Research Programme (PRIMUS/21/SCI/017); EU: H2020 European Research Council (ERC - 101002463); European Union: European Research Council (ERC - 948254, ERC 101089007), Horizon 2020 Framework Programme (MUCCA - CHIST-ERA-19-XAI-00), European Union, Future Artificial Intelligence Research (FAIR-NextGenerationEU PE00000013), Italian Center for High Performance Computing, Big Data and Quantum Computing (ICSC, NextGenerationEU); France: Agence Nationale de la Recherche (ANR-20-CE31-0013, ANR-21-CE31-0013, ANR-21-CE31-0022), Investissements d'Avenir Labex (ANR-11-LABX-0012); Germany: Baden-Württemberg Stiftung (BW Stiftung-Postdoc Eliteprogramme), Deutsche Forschungsgemeinschaft (DFG - 469666862, DFG - CR 312/5-2); Italy: Istituto Nazionale di Fisica Nucleare (ICSC, NextGenerationEU); Japan: Japan Society for the Promotion of Science (JSPS KAKENHI JP21H05085, JSPS KAKENHI JP22H01227, JSPS KAKENHI JP22H04944, JSPS KAKENHI JP22KK0227); Netherlands: Netherlands Organisation for Scientific Research (NWO Veni 2020 - VI.Veni.202.179); Norway: Research Council of Norway (RCN-314472); Poland: Polish National Agency for Academic Exchange (PPN/PPO/2020/1/00002/U/00001), Polish National Science Centre (NCN 2021/42/E/ST2/00350, NCN OPUS nr 2022/47/B/ST2/03059, NCN UMO-2019/34/E/ST2/00393, UMO-2020/37/B/ST2/01043, UMO-2021/40/C/ST2/00187, UMO-2022/47/O/ST2/00148); Slovenia: Slovenian Research Agency (ARIS grant J1-3010); Spain: BBVA Foundation (LEO22-1-603), Generalitat Valenciana (Artemisa, FEDER, IDIFEDER/2018/048), La Caixa Banking Foundation (LCF/BQ/PI20/11760025), Ministry of Science and Innovation (MCIN \& NextGenEU PCI2022-135018-2, MICIN \& FEDER PID2021-125273NB, RYC2019-028510-I, RYC2020-030254-I, RYC2021-031273-I, RYC2022-038164-I), PROMETEO and GenT Programmes Generalitat Valenciana (CIDEGENT/2019/023, CIDEGENT/2019/027); Sweden: Swedish Research Council (VR 2018-00482, VR 2022-03845, VR 2022-04683, VR grant 2021-03651), Knut and Alice Wallenberg Foundation (KAW 2017.0100, KAW 2018.0157, KAW 2018.0458, KAW 2019.0447); Switzerland: Swiss National Science Foundation (SNSF - PCEFP2\_194658); United Kingdom: Leverhulme Trust (Leverhulme Trust RPG-2020-004); United States of America: U.S. Department of Energy (ECA DE-AC02-76SF00515), Neubauer Family Foundation.


\clearpage
\printbibliography
 
\clearpage
 
\begin{flushleft}
\hypersetup{urlcolor=black}
{\Large The ATLAS Collaboration}

\bigskip

\AtlasOrcid[0000-0002-6665-4934]{G.~Aad}$^\textrm{\scriptsize 103}$,
\AtlasOrcid[0000-0001-7616-1554]{E.~Aakvaag}$^\textrm{\scriptsize 16}$,
\AtlasOrcid[0000-0002-5888-2734]{B.~Abbott}$^\textrm{\scriptsize 121}$,
\AtlasOrcid[0000-0002-1002-1652]{K.~Abeling}$^\textrm{\scriptsize 55}$,
\AtlasOrcid[0000-0001-5763-2760]{N.J.~Abicht}$^\textrm{\scriptsize 49}$,
\AtlasOrcid[0000-0002-8496-9294]{S.H.~Abidi}$^\textrm{\scriptsize 29}$,
\AtlasOrcid[0000-0002-9987-2292]{A.~Aboulhorma}$^\textrm{\scriptsize 35e}$,
\AtlasOrcid[0000-0001-5329-6640]{H.~Abramowicz}$^\textrm{\scriptsize 152}$,
\AtlasOrcid[0000-0002-1599-2896]{H.~Abreu}$^\textrm{\scriptsize 151}$,
\AtlasOrcid[0000-0003-0403-3697]{Y.~Abulaiti}$^\textrm{\scriptsize 118}$
% ...
\bigskip
\\

% ...
$^{16}$Department for Physics and Technology, University of Bergen, Bergen; Norway.\\
% ...
$^{29}$Physics Department, Brookhaven National Laboratory, Upton NY; United States of America.\\
% ...
$^{35}$$^{(a)}$Facult\'e des Sciences Ain Chock, R\'eseau Universitaire de Physique des Hautes Energies - Universit\'e Hassan II, Casablanca;$^{(b)}$Facult\'{e} des Sciences, Universit\'{e} Ibn-Tofail, K\'{e}nitra;$^{(c)}$Facult\'e des Sciences Semlalia, Universit\'e Cadi Ayyad, LPHEA-Marrakech;$^{(d)}$LPMR, Facult\'e des Sciences, Universit\'e Mohamed Premier, Oujda;$^{(e)}$Facult\'e des sciences, Universit\'e Mohammed V, Rabat;$^{(f)}$Institute of Applied Physics, Mohammed VI Polytechnic University, Ben Guerir; Morocco.\\
% ...
$^{49}$Fakult\"{a}t Physik , Technische Universit{\"a}t Dortmund, Dortmund; Germany.\\
% ...
$^{55}$II. Physikalisches Institut, Georg-August-Universit\"{a}t G\"ottingen, G\"ottingen; Germany.\\
% ...
$^{103}$CPPM, Aix-Marseille Universit\'e, CNRS/IN2P3, Marseille; France.\\
% ...
$^{118}$Department of Physics, New York University, New York NY; United States of America.\\
% ...
$^{121}$Homer L. Dodge Department of Physics and Astronomy, University of Oklahoma, Norman OK; United States of America.\\
% ...
$^{151}$Department of Physics, Technion, Israel Institute of Technology, Haifa; Israel.\\
$^{152}$Raymond and Beverly Sackler School of Physics and Astronomy, Tel Aviv University, Tel Aviv; Israel.\\
% ...
$^{a}$ Also Affiliated with an institute covered by a cooperation agreement with CERN.\\
$^{b}$ Also at An-Najah National University, Nablus; Palestine.\\
$^{c}$ Also at Borough of Manhattan Community College, City University of New York, New York NY; United States of America.\\
$^{d}$ Also at Center for High Energy Physics, Peking University; China.\\
$^{e}$ Also at Center for Interdisciplinary Research and Innovation (CIRI-AUTH), Thessaloniki; Greece.\\
$^{f}$ Also at Centro Studi e Ricerche Enrico Fermi; Italy.\\
% ...

\end{flushleft}


\clearpage

\end{document}